\documentclass[pra,aps,twocolumn,superscriptaddress,showpacs]{revtex4-1}
\usepackage{braket}
\usepackage{soul}
\usepackage{hyperref}
\usepackage{dsfont}
\usepackage{amssymb}
\usepackage{amsmath}
\usepackage{amsfonts}
\usepackage{mathtools}
\usepackage{lipsum}
\usepackage{graphicx}
\usepackage{dcolumn}
\usepackage[dvipsnames]{xcolor}
\usepackage{bm}
\usepackage{multirow}
\graphicspath{{fig/}}
\newcommand{\affilITMO}{School of Physics and Engineering, ITMO University, St. Petersburg 197101,  Russia}
\usepackage{physics}
\usepackage{appendix}

\definecolor{linkcolor}{HTML}{0176ba}
\definecolor{urlcolor}{HTML}{0176ba} 
\definecolor{citecolor}{HTML}{900020}
\hypersetup{pdfstartview=FitH,  linkcolor=linkcolor,urlcolor=urlcolor, citecolor=citecolor, colorlinks=true}

\begin{document}

\title[]{Excitation of surface plasmon-polaritons through optically-induced ultrafast transient gratings}

\author{Olesia Pashina}
  
  \affiliation{University of Brescia, Brescia, Italy
}%
 \affiliation{\affilITMO}

\author{Albert Seredin}%

\affiliation{\affilITMO 
}%

\author{Giulia Crotti}
\affiliation{Polytechnic University of Milan, Milan, Italy
}%
\author{Giuseppe Della Valle}
\affiliation{Polytechnic University of Milan, Milan, Italy
}%

\author{Andrey Bogdanov}
\affiliation{\affilITMO
}%
\affiliation{Qingdao Innovation and Development Centre, Harbin Engineering University, Qingdao, Shandong 266000, China}

\author{Mihail Petrov}
\email{m.petrov@metalab.ifmo.ru}
\affiliation{\affilITMO
}%

\author{Costantino De Angelis}
\affiliation{University of Brescia, Brescia, Italy
}%


\begin{abstract}

Ultrafast excitation of non-equilibrium carriers   under intense pulses offers unique opportunities for controlling the optical properties of semiconductor materials. In this work, we propose a scheme for ultrafast generation of surface plasmon polaritons (SPPs) via a transient metagrating formed under two interfering optical pump pulses in the semiconductor GaAs thin film. The grating can be formed due to   modulation of the refractive index associated with the non-equilibrium carrier generation. The formed temporal grating structure enables the generation of SPP waves at the GaAs/Ag interface  via  weak probe pulse excitation. We propose a theoretical model describing non-equilibrium carriers formation and diffusion and their contribution to permittivity  modulation via Drude and band-filling mechanisms. We predict that by tuning the parameters of the pump and probe one can reach the critical coupling regime and achieve efficient generation of SPP at the times scales of 0.1-1 ps.    
\end{abstract}

\maketitle





\section{Introduction}
The recent progress in nonlinear all-dielectric nanophotonics~\cite{kuznetsov2016optically, kivshar2018all, koshelev2020subwavelength,smirnova2016multipolar, grinblat2021nonlinear} naturally stimulated active studies of all-optical modulation of  semiconductor and dielectric nanostructures such as single scatterers~\cite{pogna2021ultrafast, makarov2015tuning} and metasurfaces~\cite{maiuri2024ultrafast, shcherbakov2017ultrafast, sinev2021observation,  shcherbakov2015ultrafast, zheng2023advances,gennaro2022nonlinear, yang2024ultrafast,della2017nonlinear,shilkin2024ultrafast}. On this path,  thermooptical effect, probably, shows the largest values of  Kerr-type nonlinearity leading to efficient modulation of linear and nonlinear light scattering~\cite{celebrano2021optical,rocco2021opto, tsoulos2020self, rahmani2017reversible,zhang2020anapole} and reaching bistability regimes~\cite{duh2020giant, nishida2023optical, zograf2021all,ryabov2022nonlinear}, but the modulation time scales can not be shorter than nanoseconds~\cite{husko2006ultrafast,duh2020giant}. At the same time, non-equilibrium carrier generation results in optical modulation at much shorter time scales ranging from hundreds of femtoseconds to picoseconds~\cite{di2020broadband}.  

The ultrafast nonlinearity driven by the non-equilibrium carriers has been  already utilized for Kerr-type self-induced modification of scattering of single semiconductor metaatoms~\cite{makarov2015tuning,pogna2021ultrafast} and light reflection and transmission in  metasurface structures~\cite{shcherbakov2015ultrafast, sinev2021observation,yang2024ultrafast, shilkin2024ultrafast} including fast light modulation in anisotropic  structures~\cite{della2017nonlinear}. Besides the self-induced modulation of optical properties, all-optical ultrastrong modulation of all-dielectric metasurfaces  was  achieved in the pump-probe scheme~\cite{crotti2024giant, tognazzi2023giant, yang2015nonlinear} also with help of formation of polaritonic condensates in novel excitonic materials~\cite{masharin2024giant, berte2024emergent}.

\begin{figure}[t]
    \centering    \includegraphics[width=1\columnwidth]{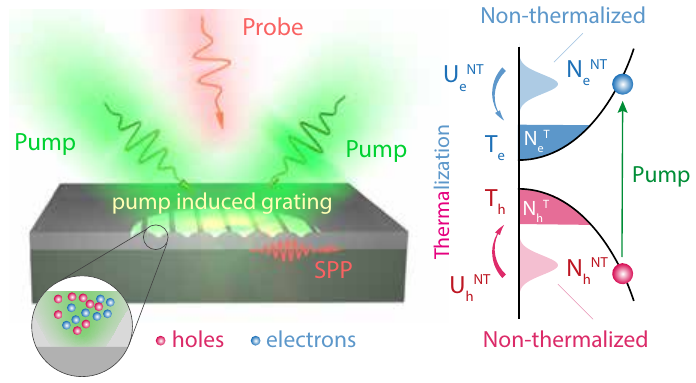}
    \caption{ Illustration of optically induced transient grating formation in GaAs film under the interference of pump fs-pulses with wavelength $\lambda_{\text{pump}}$. The grating is formed due to generation of non-equilibrium free-carriers under ultrashort pump excitation. Excitation of SPP wave at GaAs/Ag interface with probe pulse with wavelength $\lambda_{\text{pr}}$.
           }  
     \label{fig1_scheme}
\end{figure}

The streamline strategy underlying the ultrafast all-optical manipulation and control of light is based on semiconductor nanostructures, such as single nanoantennas and metasurfaces, with pronounced resonant properties~\cite{kuznetsov2016optically,staude2019all, koshelev2020subwavelength}: indeed, resonances allow to concentrate fields on subwavelength scales, granting superior modulation efficiency in compact structures with respect to non-resonant systems, e.g., thin films. On the other hand, fabrication of nanometric components with optimal quality for applications still represents a challenge for manufacturer. Moreover, in terms of reconfigurability of the optical functionality, regardless the physical mechanism exploited for the reconfiguration (being electrical, thermal, or hot-carrier based), nanostructured materials pose a further substantial issue, i.e.~the addressability of their individual nanoelements.

We present here a possible approach to tackle these limitations, based on the concept of spatial inhomogeneity in the photo-induced permittivity variation, which has been recently explored in metals~\cite{sivan2020ultrafast,Schirato2020,Schirato2022}. We exploit non-uniform modulation of the permittivity in an initially homogeneous medium to obtain the formation of a transient nanophotonic structure. 
Specifically, we theoretically propose a strategy to induce an optical grating in a thin GaAs film by illuminating it with structured light. We then show how to exploit this phenomenon to excite surface plasmon-polaritons (SPPs) at the interface between the semiconductor thin film and a metallic medium. 

The grating emergence is obtained with two crossed femtosecond pump pulses, whose interference grants a spatially periodic absorption pattern (see Fig.~\ref{fig1_scheme}). In turn, this causes a periodic modulation of the permittivity $\Delta\varepsilon$ due to the non-equilibrium carriers' distribution: thus, the grating structure is transient, and exists at the time scale of carrier diffusion and recombination. 
In this ultrafast temporal window, SPPs can be excited at the GaAs/Ag interface with a probe pulse, as the diffraction grating allows to fulfil the necessary phase-matching condition~\cite{Zheng2019Oct}.  

To quantitatively discuss the key ideas presented above, we developed a theoretical description for the non-equilibrium carriers generation and transport dynamics at the nanoscale. Our self-consistent model represents a considerable expansion and improvement with respect to the approaches employed up-to-now to study the ultrafast response of semiconductor nanostructures, photo-excited with pulsed light.

\section{Theoretical model}



 \textbf{Extended two-temperature model.} Modelling of  optically induced transient structures  requires comprehensive  analysis of free carrier generation, thermalization, diffusion, and recombination, as well as their interaction with the lattice phonons. The most common approach is based on the extended two-temperature (eTT) model~\cite{sivan2020ultrafast}, which has already been actively used for describing ultrafast dynamics in metals~\cite{sivan2020ultrafast} and metallic nanostructures~\cite{dubi2019hot}. Here we utilize eTT to model the dynamics of nonequilibrium processes occurring within a GaAs film when subjected to ultrashort laser pulse irradiation. Specifically,  generation of nonequilibrium electron and hole charge carriers are caused by optically induced interband transitions. While the detailed description of the theoretical model is provided in  Supplemental Information, Section 1, here we will present the key aspects of the model.

The formed non-thermalized carriers are described by the concentration  ${N}_{\mathrm{ e}}^{\mathrm{NT}}(\mathbf{r}, t)$ and corresponding energy $\mathcal{U}_{\mathrm{e}}^{ \mathrm{NT}}( \mathbf{r}, t)$ as shown in see Fig.~\ref{fig1_scheme}. The Dember effect~\cite{oguri2015dynamical}, which facilitates ambipolar diffusion of generated electrons and holes, enables us to assume that the evolution of charge carriers occurs in a nearly identical manner. Consequently, we can analyze our problem exclusively from the perspective of single charges, i.e. electrons. 
The subsequent carriers thermalization  results in formation of their quasi-equilibrium distribution  with temperature $T_e(\mathbf{r}, t)$ and concentration  ${N}_{\mathrm{ e}}^{\mathrm{T}}(\mathbf{r}, t)$.  The further cooling of electron and hole subsystem due to interaction with phonons leads to increase of the lattice temperature  $T_{ph}(\mathbf{r}, t)$. The model also accounts for non-thermalized and thermalized carrier diffusion and relaxation, as well as for heat generation and diffusion. One can see the characteristic times of mentioned processes occurring in the structure in Table~\ref{table2}.

The approach is based on the system of coupled differential equations describing the key characteristics dynamics with a spatiotemporal resolution (below each characteristic is assumed with $(\mathbf{r}, t)$ dependence):

\begin{equation}
\begin{gathered}
\frac{\partial \mathcal{U}_{\mathrm{e}}^{\mathrm{NT}}}{\partial t}=\frac{1}{C_{\mathrm{e}}} \nabla\left[k_{\mathrm{e}}\nabla \mathcal{U}_{\mathrm{e}}^{\mathrm{NT}}\right]-\left(\Gamma_{\mathrm{e}}+\Gamma_{\mathrm{ph}}\right) \mathcal{U}_{\mathrm{e}}^{\mathrm{NT}}+Q_{\mathcal{U}_{\mathrm{e}}}, \\
\frac{\partial N_e^{\mathrm{NT}}}{\partial t}=Q_{{N_e}}+\nabla\left[D \nabla N_e^{\mathrm{NT}}\right]-\Gamma_{\mathrm{e}} N_e^{\mathrm{NT}},\\
\frac{\partial N_e^{\mathrm{T}}}{\partial t}=
\Gamma_e N_e^{\mathrm{NT}}
+ \nabla \left[D \nabla N_e^{\mathrm{T}}\right]-\\(\gamma_{nr} {N_e^{\mathrm{T}}+\gamma_{r} {N_e^{\mathrm{T}}}^{2}+\gamma_{Au} {N_e^{\mathrm{T}}}^{3})
},\\
\frac{\partial [C_e  T_e]}{\partial t} = \nabla \left[k_e \nabla T_e\right]  - G_{\mathrm{e}-\mathrm{ph}}[T_e-T_{ph}] +\Gamma_{\mathrm{e}} \mathcal{U}_{\mathrm{e}}^{\mathrm{NT}},\\
\frac{\partial [C_{ph}T_{ph}]}{\partial t} = \nabla[k_{ph} \nabla T_{ph}] + G_{\mathrm{e}-\mathrm{ph}}[T_e-T_{ph}] +\Gamma_{\mathrm{ph}}\mathcal{U}_{\mathrm{e}}^{\mathrm{NT}},
\end{gathered}
\label{eq_syst}
\end{equation}
Here $Q_{{N_e}}$ and $Q_{{\mathcal{U}_e}}$ represent the concentration and energy sources characterized by optical absorption at each point of the structure defined by the imaginary part of the dielectric permittivity $\varepsilon(\mathbf{r},t)$. The coefficients $C_{e/ph},\: k_{e/ph}, \:\mu_{e/ph}$  correspond to the  electron/phonon heat capacity, thermal conductivity, and mobility~\cite{Schumann1991Jan, Derrien2011, BibEntry2001Jul}, respectively. Additionally, $\Gamma_e$ denotes the rate of energy exchange between non-thermal and thermal electrons~\cite{Hopfel1988Apr}, while $\Gamma_{\mathrm ph}$ is the rate of energy transfer between electrons and phonons~\cite{Sivan2020May}. The electron-phonon relaxation time is given by $G_{\mathrm{e}-\mathrm{ph}}$~\cite{Sivan2020May, Margiolakis2018Dec}, and $D$  is the coefficient for ambipolar heat diffusion~\cite{Ruzicka2010Dec}.
Model also accounts for
non-radiative decay $\gamma_{nr}$, radiative recombination $\gamma_{r}$ and Auger recombination $\gamma_{Au}$ in GaAs~\cite{Strauss1993Jan}. 

\begin{table}[h]
    \caption{ The characteristic time scales of various ultrafast processes underlying the two-temperature model of the GaAs material sorted in ascending order. These values are defined for electron temperatures $T_e=300-10^4~\mathrm{K}$ and high-density electron-hole plasma $N_{e} = 10^{20}~\mathrm {cm^{-3}}$ taking into account the characteristic nonuniformity length scale $d \approx 1~{\mu \mathrm{m}}$. Here, electron-phonon scattering is related to $\Gamma^{T_{e}}(N_e)=G_{\mathrm{e}-\mathrm{ph}}(N_e)/C_{e}(N_e)$ and $\Gamma^{T_{ph}}(N_e, T_{ph})=G_{\mathrm{e}-\mathrm{ph}}(N_e)/C_{ph}(T_{ph})$. The diffusion of electron concentration can be characterized by the concentration diffusion time $\tau_\mathrm{diff}^{N_{e}} = \frac{d^2}{4\pi^2D}$, while the heat diffusion coefficients for electrons and phonons are defined as $\tau_\mathrm{diff}^{T_{e/ph}} = \frac{d^2}{4\pi^2D_{e/ph}}$, where $D_{e/ph}=k_{e/ph}/C_{e/ph}$~\cite{block2019tracking}(See more details in Supplemental Information Sec. 1).}
        \label{table2}
\centering

\begin{tabular}{| m{4cm} | m{2.2cm}| m{2.1cm}| } 
 \hline   \centering
\footnotesize{\textbf{Effect}}  &   \centering\footnotesize{\textbf{Characteristic time scale}} &  \quad \footnotesize{ \textbf{Value}}\\
 \hline
  \centering
\footnotesize{Thermalization of non-thermal electrons $\mathcal{U}_{\mathrm{e}}^{\mathrm{NT}}$} & \centering \footnotesize{$\tau_{e}=1/\Gamma_{e}(T_e)$}  & \footnotesize{$\approx 200$ fs} \\
  \hline
    \centering
   \footnotesize{Evolution of electron temperature $T_e$ due to electron-phonon scattering}  & \centering \footnotesize{$1/\Gamma^{T_{e}}(N_e)$}  & \footnotesize{$\approx 500$ fs} \\
  \hline
    \rule{0cm}{0.45cm}
  \centering \footnotesize{ Diffusion of electron temperature $T_e$} & \centering \footnotesize{$\tau_\mathrm{diff}^{T_e}(d, T_e)$ } & \footnotesize{$\approx2$ ps - 52 fs} \\ 
  \hline
      \rule{0cm}{0.45cm}
  \centering \footnotesize{Diffusion of electron concentration $N_e$}  & \centering \footnotesize{$\tau_\mathrm{diff}^{N_e}(d, T_e)$ } &   \footnotesize{ $\approx13$ ps - 390 fs} \\
 \hline
   \centering
\footnotesize{Evolution of phonon temperature $T_{ph}$ due to electron-phonon scattering } & \centering \footnotesize{$1/\Gamma^{T_{ph}}(N_e)$}  & \footnotesize{$\approx 0.3$ ns} \\
  \hline
  \centering
  \footnotesize{Energy transfer between non-thermal electrons and phonons} & \centering \footnotesize{$1/\Gamma_{ph}(N_e)$ } & \footnotesize{$\approx 0.3$ ns}\\
  \hline
 \rule{0cm}{0.45cm}
 \centering
  \footnotesize{ Diffusion of phonon temperature $T_{ph}$ }& \centering \footnotesize{ $\tau_\mathrm{diff}^{T_{ph}}(d)$}  & \footnotesize{$\approx 80$ ns}\\ 
  \hline

    \end{tabular}
\end{table}

All coefficients in the system are dependent on the evolving  parameters, which ultimately makes this approach complex and highly self-consistent. The self-consistency is also caused by the temporal modulation of dielectric permittivity and optical properties of structures, driven by the dynamics of carriers and phonons characteristics, which in terms changes the source functions  $Q_{{N_e}}$ and $Q_{{\mathcal{U}_e}}$.

 \textbf{Dielectric permittivity.} We account altering of the  dielectric constant of the material via two main effects: formation of free-electron gas accounted within the  Drude model~\cite{Sokolowski-Tinten2000Jan}, and band-filling effect~\cite{Bennett1990Feb} related to the occupation of the states at the bottom of the conductance band (top of the valence band) with the free carriers (see more details in Supplemental Information Sec. 2). Thus, the  dielectric constant of has two contributions associated with the Drude model and band-filling effect: 
\begin{align}
\varepsilon_\mathrm{fin}(\mathbf{r},t) = [(n_0+&\Delta n_{bf}(\mathbf{r},t)) +i(k_0+\Delta k_{bf})(\mathbf{r},t)]^2 +...\nonumber\\
&+\Delta\varepsilon_{Dr}(\mathbf{r},t),
\end{align} where $n_0 + ik_0$ is the initial complex refractive index
of GaAs~\cite{Papatryfonos2021Feb}, while the terms $\Delta n_{bf}(\mathbf{r},t)+i\Delta k_{bf}(\mathbf{r},t)$ and 
$\Delta\varepsilon_{Dr} (\mathbf{r},t)$ represent the change of complex refractive index due to the band filling effect and the complex dielectric permittivity modulation based on Drude model, respectively.
Within the model, we also assume that dielectric constant $\varepsilon(\textbf{r}, t)$ varies in time slowly enough  that we can consider the problem segregated manner and solving stationary Maxwell's equations at each time moment $t$. This approximation is valid for pulses of hundreds of femtosecond and  optical systems with relatively low Q-factor,  and self-consistent solution of Maxwell's equations  along with a carrier dynamics equations may be required. However, our research indicates that accounting for this self-consistency in the context of ultrashort pulses (with femtosecond duration) is negligible, but for longer pulse duration the developed model is essential (See Sec. 4B in SI).

Finally, we implemented our theoretical model within the COMSOL Multiphysics simulation package (See Sec. 3 in SI) by linking Maxwell equations with diffusion equations describing carriers dynamics in a  self-consistent manner. This allows us to explore the connection between the electromagnetic field distribution $\mathbf{E}(\mathbf{r}, t)$ within the semiconductor material and the evolution of the key parameters ($\mathcal{U}_{\mathrm{e}}^{ \mathrm{NT}}( \mathbf{r}, t),{N}_{\mathrm{ e}}^{\mathrm{NT}}(\mathbf{r}, t),{N}_{\mathrm{ e}}^{\mathrm{T}}(\mathbf{r}, t),{T}_{\mathrm{ e}}(\mathbf{r}, t),{T}_{\mathrm{ph}}(\mathbf{r}, t)$) resulting in the overall variation of the optical constants and change in optical properties of the photonic structure.

 \begin{figure}[h!]
    \centering
    \includegraphics[width=0.8\columnwidth]{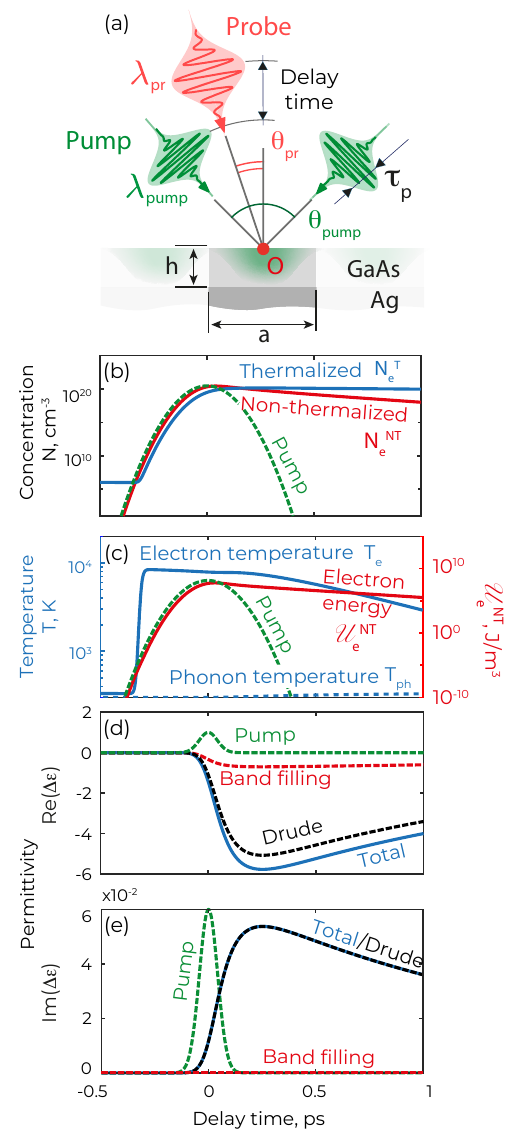}
    \caption{(a) Interference of two pump beams forming a lattice with period $a$ in GaAs film of height $h$.  The dynamics of non-thermalized $N_e^\mathrm{NT}$  (b) and thermalized  $N_e^\mathrm{T}$ (c) electrons concentration, energy of non-thermalized electrons $\mathcal{U}_e^\mathrm{NT}$ and electron temperature $T_{e}$ (d) and phonon temperature $T_{ph}$ computed at the  point O in (a).  b) Evolution of real (d) and imaginary (e) parts of the dielectric permittivity correspondent to the wavelength 1600 nm and computed at point O.  Drude model (black dashed line) and band filling model (red dashed line) contributions  are shown along with their sum ('total', blue solid line). The Gaussian pump profile is shown with a dashed green line.  }  
     \label{fig2_evolution}
\end{figure} 


\section{Transient grating formation in \text{GaAs} film}

We applied developed theoretical model and numerical approach to describe formation of a transient diffraction grating within a GaAs film with a height of $h=300$ nm on an Ag substrate. To achieve this, we excited the film using two 'pump' plane wave  pulses with  central wavelength $\lambda_\mathrm{pump} =500$ nm impinging the surface at angles $\theta_\mathrm{pump} =\pm13 ^\mathrm{o}$ as shown in Fig.~\ref{fig2_evolution} (a). The pulse has Gaussian profile  $ \mathrm{GP}(t) =\mathrm{I_{{pump}}} \mathrm{exp} (-4 \cdot\mathrm{ln}2 \cdot t^2/\tau_\mathrm{p}^2)$ reaching the maximal intensity about $\mathrm{I_{pump}}= 10^{14}~\mathrm{W/m^2}$ that leads to significant evolution of the optical parameters of the film, while the pulse duration  is set to $\tau_{\mathrm{p}}=100~$fs. These pump intensity and pulse duration correspond to a fluence of $\mathrm{F} = 1~\mathrm{mJ/cm^2}$, which is below the damage threshold in GaAs~\cite{di2020broadband,glezer1995laser,singh2002ripples}. The resulting periodical interference pattern triggers intense generation of non-equilibrium carriers. The period of this interference is $\mathrm{a}=1100$ nm  corresponding to the angle of pump incidence of approximately $\mathrm{\theta}_{\mathrm{pump}} \approx \arcsin{({\lambda_\mathrm{pump}}/(2\mathrm{a}))}$.


In Fig.~\ref{fig2_evolution}, we plot the evolution of non-equilibrium carrier dynamics computed at one point close to the surface (point O in Fig.~\ref{fig2_evolution} (a)). At first, the absorbed energy of the pump pulse is transferred to the non-thermalized carriers with the rapid growth of their concentration $N_e^{NT}$ (up to the values about $10^{21}$ cm$^{-3}$, see red solid line in Fig.~\ref{fig2_evolution}(b)) and energy $\mathcal{U}_{\mathrm{e}}^{ \mathrm{NT}}$ (see red solid line in Fig.~\ref{fig2_evolution}(c)). After that, at the picosecond timescale, the carriers are being thermalized giving rise to $N_e^{T}$ concentration and carrier temperature $T_e$ (blue solid lines in Fig.~\ref{fig2_evolution}(b) and (c) correspondingly). The carrier temperature is increased up to the values of $8\cdot 10^4-9\cdot 10^4$ K and then gradually decreases after the end of the pump pulse due to electron-phonon interactions. At the same time, the phonon temperature increases for less than hundred degrees at this timescale (see blue dashed line in Fig.~\ref{fig2_evolution}(c)). A more detailed analysis can be found in Sec. 4A in SI.

Finally, the predicted increase of free carrier concentration results in a drastic change of  the complex dielectric permittivity $\mathrm{Re ~\varepsilon_\mathrm{pr}}+i\mathrm{Im ~\varepsilon_\mathrm{pr}}$ at the probe wavelength $\lambda_{\text{pr}}=1600$ nm as shown in FigS.~\ref{fig2_evolution}(d) and ~\ref{fig2_evolution}(e). 
At the first stage,  non-thermalized electrons are generated immediately after the pump pulse arrival. These electrons subsequently undergo thermalization and diffusion, leading to a corresponding modulation of the dielectric constant. One can see that at the current pump intensity, the real part of the permittivity can be reduced for 4-6 units at  time scale of 1 ps which constitute  almost 50\% of the non-perturbed dielectric constant $\varepsilon^{0} = 11.4 +i0$~\cite{Papatryfonos2021Feb}. This variation is predominantly provided by the Drude contribution, while the band filling component is relatively small. The increase in the imaginary part of the permittivity is clearly related to generation of free-carriers, {as the impact of the band filling effect is zero for the probe wavelength of 1600 nm, which is higher than the band gap wavelength.} At the same time, above the band gap, at the pump wavelength $\lambda_{\text{pump}}=500$ nm for instance, both the Drude model and the band filling effect  play a substantial role in modulating the dielectric constant of the material (see Sec.~2C in SI). 

\begin{figure}[t]
    \centering
    \includegraphics[width=\columnwidth]{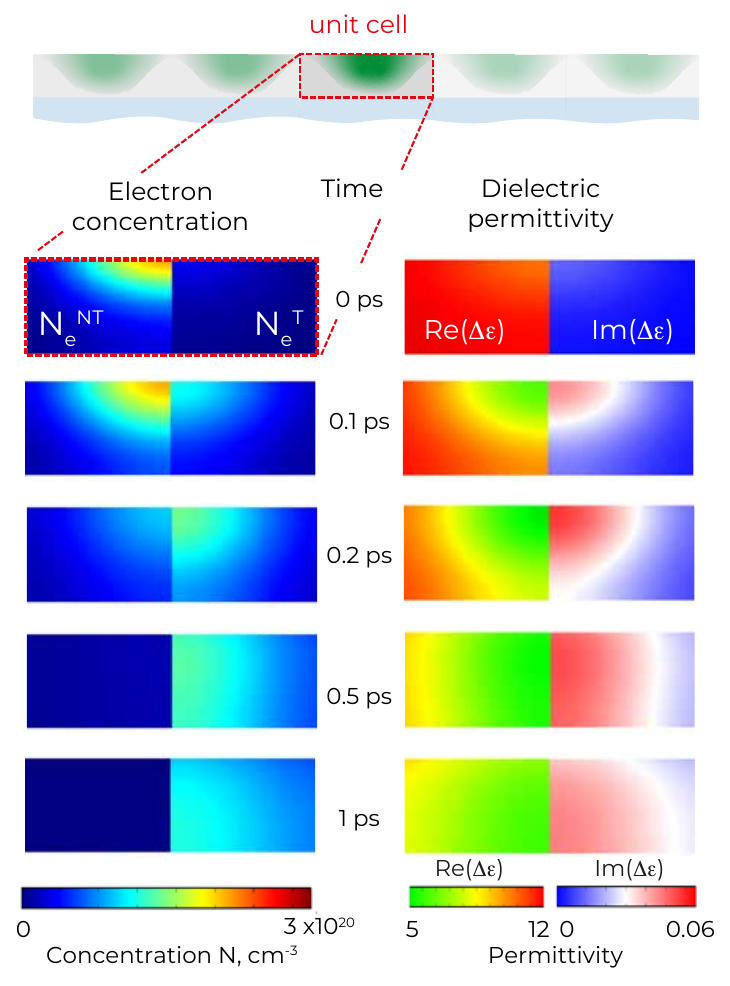}
    \caption{The distribution of thermalized and non-thermalized electrons concentrations (left column) and real and imaginary  parts of  dieletric permittivity (right column) for probe wavelength $\lambda_{\mathrm{pr}}= 1600$ nm within the unit cell in GaAs film at different delay time.}      
     \label{fig3_spatial_evolution}
\end{figure} 


\begin{figure*}[t]
    \centering
    \includegraphics[width=2\columnwidth]{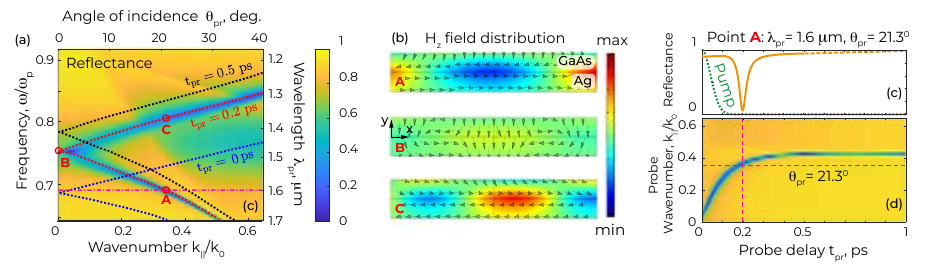}
    \caption{(a) Reflectance map for the probe irradiation as a function of probe wavelength $\lambda_{\mathrm{pr}}$  and angle of incidence $\theta_{\mathrm{pr}}$ (frequency $\omega/\omega_{p} = a/\lambda_{\mathrm{pr}}$ and lateral wavenumber $k_{||}/k_0 = \sin (\theta_{\mathrm{pr}})$). The map corresponds to a fixed probe time $\mathrm{t_{pr}} = 0.2$ ps. The evolution of SPP dispersion over probe time is demonstrated through dashed lines for probe delay time $\mathrm{t_{pr}} =0, 0.2, 0.5$ ps. 
    (b) The magnetic field distribution $\mathrm{H_z}$ near the GaAs/Ag interface at a probe delay time $\mathrm{t_{pr}} = 0.2$ ps for three specific excitation points labeled as A, B, and C in Fig.~\ref{fig4_SPP} (a)). The arrows indicate the direction of Poynting vector. (c) The reflectance dynamics for point A ($\lambda_{\mathrm{pr}} = 1600$ nm and $\mathrm{q}_{\mathrm{pr}} = 21.3^{o}$).  The Gaussian pump profile is shown with a dashed green line.    (d) Reflectance map of the probe beam as a function of probe delay time $\mathrm{t_{pr}}$ and normalized wavenumber $k_{||}/k_0$ for fixed probe wavelength $\lambda_{\mathrm{pr}}= 1600$ nm.}
    \label{fig4_SPP}
\end{figure*}

While previous simulations show the evolution of dielectric permittivity in time at the given point close to the film interface (point O in Fig.~\ref{fig2_evolution} (a)), it is quite insightful to trace the spatial distribution of generated free carriers and correspondent variation of the dielectric permittivity.   The left column in Fig.~\ref{fig3_spatial_evolution} depicts  the spatial distribution of the free carriers in a unit cell of the film at different time delays from 0 to 1 ps. One can that the intialy formed non-thermalized electrons are fully converted to thermalized electrons within 0.2 ps at the depth of around  $h/2=150~\mathrm{nm}$, since the thermalization time $\tau_e\approx 200~\mathrm{fs}$ (See Table~\ref{table2}). The thermalized electrons undergo the following diffusion in the depth of the film. The variation of the real and imaginary part of the dielectric permittivity (right column in Fig.~\ref{fig3_spatial_evolution}) generally follows the free-carriers concentration forming an optical grating with high contrast. One can also notice that already at 1 ps delay the optical grating is almost vanished. Indeed, the grating contrast if defined by the inhomogeneity of the hot carriers distribution, at the same time within 1 ps   the hot carriers become homogeneously distributed across the film driven by diffusion process (See Table~\ref{table2}). Note, that the dielectric permittivity of the film is far from the equilibrium ($\Re \varepsilon\approx5$) and it will be kept at this values on the timescale of free carrier recombination, around 5-10 ns. Thus, the lifetime of grating is much faster than the life time of free carriers and is defined by the carrier diffusion time $\tau_\mathrm{diff}^{N_e}$ (See Table~\ref{table2}).

\section{Excitation of SPPs}

The  optically-induced  diffraction grating formed within the GaAs film at picosecond timescale enables the ultrafast excitation of SPPs at the  semiconductor/metal interface. As it is widely recognized, SPPs are tightly confined TM electromagnetic surface waves propagating along a metal-semiconductor interface with spectrum lying below the light cone~\cite{Zheng2019Oct}. Therefore, the induced diffraction grating ensures the quasi-phasematching condition and  fulfills the quasi-wavevector conservation law, necessary for exciting SPPs from free space~\cite{Liang2015Nov} as shown in Fig.~\ref{fig1_scheme}:

\begin{equation}
\frac{2\pi}{a}+k_\mathrm{pr}\sin(\theta_\mathrm{pr})=k_{SPP}(\omega_\mathrm{pr}).
\label{eq:SPP}
\end{equation}

\noindent Here, $k_\mathrm{pr}=\omega_\mathrm{pr}/c$ is the wavevector of the incident probe,  $\omega_\mathrm{pr}$ is the probe frequency, $c$ is the speed of light, $k_{SPP}(\omega_\mathrm{pr})$ is  the dispersion relation of SPP at the probe frequency.

The excitation of SPPs can be observed in the reflectance spectra of the probe beam. For that we plotted the reflectance amplitude as function of the probe wavelength and incidence angle shown in Fig.~\ref{fig4_SPP} (a) at time delay of $t_{\text{pr}}=0.2$ ps after the pump pulse arrival. One can see a distinct dip in the reflectance amplitude corresponding to the excitation of SPP. The SPP dispersion according to Eq.\eqref{eq:SPP} is shown with a red dashed line for the time delay $t_{\text{pr}}=0.2$ ps. This time corresponds to the point of maximum variation in the real part of the dielectric permittivity $\Delta \mathrm{Re~\varepsilon_{pr}}$, as depicted in Fig.~\ref{fig3_spatial_evolution}. 

The distribution of $\mathrm{H_z}$ field component for a fixed probe wavelength  $\lambda_{\mathrm{pr}} = 1600$ nm and angle $\mathrm{\theta}_{\mathrm{pr}} = 21.3^{\mathrm{o}}$ related to the point A in Fig.~\ref{fig4_SPP}(a) is shown in Fig.~\ref{fig4_SPP}(b) with a clear signature of SPP mode. The temporal evolution of the reflectance amplitude for the same parameters of the probe wavelength and angle is shown in Fig.~\ref{fig4_SPP}(c) demonstrating ultrafast excitation of SPP with FWHM about 40 femtoseconds reaching almost zero reflectance. It is worth mentioning that nearly total reduction in reflectance was achieved by adjusting the film thickness of the GaAs film $h$ under fixed pump and probe irradiation conditions ensuring the critical coupling regime. Additionally, at the same angle of incidence $\theta_{\mathrm{pr}} = 21.3^{o}$ and probe wavelength $\lambda_{\mathrm{pr}} = 1360$ nm (point C) it is feasible to achieve SPP excitation propagating in the opposite direction along the metal-semiconductor interface. The reverse propagation of SPP  is indicated by the Pointing vector orientation depicted by arrows in Fig.~\ref{fig4_SPP}(b)  illustrating the energy flow along the interface of the structure. This figure also illustrates the magnetic field distribution $\mathrm{H_{z}}$ at  normal incidence of the probing field (point B). The latter relates to the excitation of two oppositely directed SPPs, leading to the formation of a standing SPP wave.

Finally, in Fig.~\ref{fig4_SPP}(d), we observe that the evolution of the diffraction grating, driven by the diffusion and recombination of induced nonequilibrium carriers in the conduction band (Fig.~\ref{fig3_spatial_evolution}), causes a shift in the dispersion of the surface plasmon polariton towards shorter wavelength (see in more detail in Supplemental Information Sec. 4C). 

This phenomenon opens up an unprecedented way for the ultrafast control over SPPs. Actually, when operating at a fixed wavelength $\lambda_{\mathrm{pr}}$ (for example, 1600 nm) and suitable angle of incidence $\mathrm{\theta}_\mathrm{pr}$ ($21.3^{\mathrm{o}}$), the phase matching with SPP is only permitted over the extremely short temporal window at around the intersection between horizontal and vertical dashed lines in Fig.~\ref{fig4_SPP}(d).


\section{Conclusion}

 We theoretically proposed  a setup for the ultrafast generation of SPP waves at semiconductor film/metal interface by  optically induced grating structure. The grating formed in a thin film of GaAs material is induced by the non-equilibrium carriers and exists at the timescales of carriers diffusion that opens an opportunity for ultrafast control over SPP below the picosecond time scale. Our results are based on a developed extended two-temperature self-consistent model that describes the ultrafast optically induced processes in GaAs film and semiconductors in general. In particular, it demonstrates that the optimal delay between the pump pulse generating the grating and the probe pulse exciting the SPP wave, as well as the optimal angle of probe pulse incidence, are fully defined by the carriers' dynamics and can be chosen in a wide range. Finally, we believe that our advanced theoretical framework concerning ultrafast semiconductor responses to pulsed electromagnetic radiation and proposed novel technique for ultrafast SSP control exhibit significant potential for diverse applications in all-optically controlled ultrafast devices.

\section{Acknowledgement}

The authors are thankful Yonatan Sivan and Zarina Sadrieva for fruitful discussions. The work was supported by the Federal Academic Leadership Program Priority 2030.  


\begin{thebibliography}{10}

\bibitem{kuznetsov2016optically}
A.~I. Kuznetsov, A.~E. Miroshnichenko, M.~L. Brongersma, Y.~S. Kivshar, and B.~Luk'yanchuk, ``Optically resonant dielectric nanostructures,'' {\em Science}, vol.~354, no.~6314, p.~aag2472, 2016.

\bibitem{kivshar2018all}
Y.~Kivshar, ``All-dielectric meta-optics and non-linear nanophotonics,'' {\em National Science Review}, vol.~5, no.~2, pp.~144--158, 2018.

\bibitem{koshelev2020subwavelength}
K.~Koshelev, S.~Kruk, E.~Melik-Gaykazyan, J.-H. Choi, A.~Bogdanov, H.-G. Park, and Y.~Kivshar, ``Subwavelength dielectric resonators for nonlinear nanophotonics,'' {\em Science}, vol.~367, no.~6475, pp.~288--292, 2020.

\bibitem{smirnova2016multipolar}
D.~Smirnova and Y.~S. Kivshar, ``Multipolar nonlinear nanophotonics,'' {\em Optica}, vol.~3, no.~11, pp.~1241--1255, 2016.

\bibitem{grinblat2021nonlinear}
G.~Grinblat, ``Nonlinear dielectric nanoantennas and metasurfaces: frequency conversion and wavefront control,'' {\em ACS Photonics}, vol.~8, no.~12, pp.~3406--3432, 2021.

\bibitem{pogna2021ultrafast}
E.~A.~A. Pogna, M.~Celebrano, A.~Mazzanti, L.~Ghirardini, L.~Carletti, G.~Marino, A.~Schirato, D.~Viola, P.~Laporta, C.~De~Angelis, {\em et~al.}, ``Ultrafast, all optically reconfigurable, nonlinear nanoantenna,'' {\em ACS nano}, vol.~15, no.~7, pp.~11150--11157, 2021.

\bibitem{makarov2015tuning}
S.~Makarov, S.~Kudryashov, I.~Mukhin, A.~Mozharov, V.~Milichko, A.~Krasnok, and P.~Belov, ``Tuning of magnetic optical response in a dielectric nanoparticle by ultrafast photoexcitation of dense electron--hole plasma,'' {\em Nano letters}, vol.~15, no.~9, pp.~6187--6192, 2015.

\bibitem{maiuri2024ultrafast}
M.~Maiuri, A.~Schirato, G.~Cerullo, and G.~Della~Valle, ``Ultrafast all-optical metasurfaces: challenges and new frontiers,'' {\em ACS Photonics}, vol.~11, no.~8, pp.~2888--2905, 2024.

\bibitem{shcherbakov2017ultrafast}
M.~R. Shcherbakov, S.~Liu, V.~V. Zubyuk, A.~Vaskin, P.~P. Vabishchevich, G.~Keeler, T.~Pertsch, T.~V. Dolgova, I.~Staude, I.~Brener, {\em et~al.}, ``Ultrafast all-optical tuning of direct-gap semiconductor metasurfaces,'' {\em Nature communications}, vol.~8, no.~1, pp.~1--6, 2017.

\bibitem{sinev2021observation}
I.~S. Sinev, K.~Koshelev, Z.~Liu, A.~Rudenko, K.~Ladutenko, A.~Shcherbakov, Z.~Sadrieva, M.~Baranov, T.~Itina, J.~Liu, {\em et~al.}, ``Observation of ultrafast self-action effects in quasi-bic resonant metasurfaces,'' {\em Nano Letters}, vol.~21, no.~20, pp.~8848--8855, 2021.

\bibitem{shcherbakov2015ultrafast}
M.~R. Shcherbakov, P.~P. Vabishchevich, A.~S. Shorokhov, K.~E. Chong, D.-Y. Choi, I.~Staude, A.~E. Miroshnichenko, D.~N. Neshev, A.~A. Fedyanin, and Y.~S. Kivshar, ``Ultrafast all-optical switching with magnetic resonances in nonlinear dielectric nanostructures,'' {\em Nano letters}, vol.~15, no.~10, pp.~6985--6990, 2015.

\bibitem{zheng2023advances}
Z.~Zheng, D.~Rocco, H.~Ren, O.~Sergaeva, Y.~Zhang, K.~B. Whaley, C.~Ying, D.~de~Ceglia, C.~De-Angelis, M.~Rahmani, {\em et~al.}, ``Advances in nonlinear metasurfaces for imaging, quantum, and sensing applications,'' {\em Nanophotonics}, vol.~12, no.~23, pp.~4255--4281, 2023.

\bibitem{gennaro2022nonlinear}
S.~Gennaro, R.~Sarma, and I.~Brener, ``Nonlinear and ultrafast all-dielectric metasurfaces at the center for integrated nanotechnologies,'' {\em Nanotechnology}, vol.~33, no.~40, p.~402001, 2022.

\bibitem{yang2024ultrafast}
Z.~Yang, M.~Liu, D.~Smirnova, A.~Komar, M.~Shcherbakov, T.~Pertsch, and D.~Neshev, ``Ultrafast q-boosting in semiconductor metasurfaces,'' {\em Nanophotonics}, vol.~13, no.~12, pp.~2173--2182, 2024.

\bibitem{della2017nonlinear}
G.~Della~Valle, B.~Hopkins, L.~Ganzer, T.~Stoll, M.~Rahmani, S.~Longhi, Y.~S. Kivshar, C.~De~Angelis, D.~N. Neshev, and G.~Cerullo, ``Nonlinear anisotropic dielectric metasurfaces for ultrafast nanophotonics,'' {\em ACS Photonics}, vol.~4, no.~9, pp.~2129--2136, 2017.

\bibitem{shilkin2024ultrafast}
D.~A. Shilkin, S.~T. Ha, R.~Paniagua-Dom{\'\i}nguez, and A.~I. Kuznetsov, ``Ultrafast modulation of a nonlocal semiconductor metasurface under spatially selective optical pumping,'' {\em Nano Letters}, 2024.

\bibitem{celebrano2021optical}
M.~Celebrano, D.~Rocco, M.~Gandolfi, A.~Zilli, F.~Rusconi, A.~Tognazzi, A.~Mazzanti, L.~Ghirardini, E.~A. Pogna, L.~Carletti, {\em et~al.}, ``Optical tuning of dielectric nanoantennas for thermo-optically reconfigurable nonlinear metasurfaces,'' {\em Optics Letters}, vol.~46, no.~10, pp.~2453--2456, 2021.

\bibitem{rocco2021opto}
D.~Rocco, M.~Gandolfi, A.~Tognazzi, O.~Pashina, G.~Zograf, K.~Frizyuk, C.~Gigli, G.~Leo, S.~Makarov, M.~Petrov, {\em et~al.}, ``Opto-thermally controlled beam steering in nonlinear all-dielectric metastructures,'' {\em Optics Express}, vol.~29, no.~23, pp.~37128--37139, 2021.

\bibitem{tsoulos2020self}
T.~V. Tsoulos and G.~Tagliabue, ``Self-induced thermo-optical effects in silicon and germanium dielectric nanoresonators,'' {\em Nanophotonics}, vol.~9, no.~12, pp.~3849--3861, 2020.

\bibitem{rahmani2017reversible}
M.~Rahmani, L.~Xu, A.~E. Miroshnichenko, A.~Komar, R.~Camacho-Morales, H.~Chen, Y.~Z{\'a}rate, S.~Kruk, G.~Zhang, D.~N. Neshev, {\em et~al.}, ``Reversible thermal tuning of all-dielectric metasurfaces,'' {\em Advanced Functional Materials}, vol.~27, no.~31, p.~1700580, 2017.

\bibitem{zhang2020anapole}
T.~Zhang, Y.~Che, K.~Chen, J.~Xu, Y.~Xu, T.~Wen, G.~Lu, X.~Liu, B.~Wang, X.~Xu, {\em et~al.}, ``Anapole mediated giant photothermal nonlinearity in nanostructured silicon,'' {\em Nature communications}, vol.~11, no.~1, p.~3027, 2020.

\bibitem{duh2020giant}
Y.-S. Duh, Y.~Nagasaki, Y.-L. Tang, P.-H. Wu, H.-Y. Cheng, T.-H. Yen, H.-X. Ding, K.~Nishida, I.~Hotta, J.-H. Yang, {\em et~al.}, ``Giant photothermal nonlinearity in a single silicon nanostructure,'' {\em Nature communications}, vol.~11, no.~1, p.~4101, 2020.

\bibitem{nishida2023optical}
K.~Nishida, P.-H. Tseng, Y.-C. Chen, P.-H. Wu, C.-Y. Yang, J.-H. Yang, W.-R. Chen, O.~Pashina, M.~I. Petrov, K.-P. Chen, {\em et~al.}, ``Optical bistability in nanosilicon with record low q-factor,'' {\em Nano Letters}, vol.~23, no.~24, pp.~11727--11733, 2023.

\bibitem{zograf2021all}
G.~P. Zograf, M.~I. Petrov, S.~V. Makarov, and Y.~S. Kivshar, ``All-dielectric thermonanophotonics,'' {\em Advances in Optics and Photonics}, vol.~13, no.~3, pp.~643--702, 2021.

\bibitem{ryabov2022nonlinear}
D.~Ryabov, O.~Pashina, G.~Zograf, S.~Makarov, and M.~Petrov, ``Nonlinear optical heating of all-dielectric super-cavity: efficient light-to-heat conversion through giant thermorefractive bistability,'' {\em Nanophotonics}, vol.~11, no.~17, pp.~3981--3991, 2022.

\bibitem{husko2006ultrafast}
C.~Husko and C.~W. Wong, ``Ultrafast all-optical bistability in algaas photonic crystals,'' in {\em Nanophotonics for Communication: Materials, Devices, and Systems III}, vol.~6393, pp.~111--117, SPIE, 2006.

\bibitem{di2020broadband}
A.~Di~Cicco, G.~Polzoni, R.~Gunnella, A.~Trapananti, M.~Minicucci, S.~Rezvani, D.~Catone, L.~Di~Mario, J.~Pelli~Cresi, S.~Turchini, {\em et~al.}, ``Broadband optical ultrafast reflectivity of si, ge and gaas,'' {\em Scientific Reports}, vol.~10, no.~1, p.~17363, 2020.

\bibitem{crotti2024giant}
G.~Crotti, M.~Akturk, A.~Schirato, V.~Vinel, A.~A. Trifonov, I.~C. Buchvarov, D.~N. Neshev, R.~Proietti~Zaccaria, P.~Laporta, A.~Lema{\^\i}tre, {\em et~al.}, ``Giant ultrafast dichroism and birefringence with active nonlocal metasurfaces,'' {\em Light: Science \& Applications}, vol.~13, no.~1, p.~204, 2024.

\bibitem{tognazzi2023giant}
A.~Tognazzi, P.~Franceschini, O.~Sergaeva, L.~Carletti, I.~Alessandri, G.~Finco, O.~Takayama, R.~Malureanu, A.~V. Lavrinenko, A.~C. Cino, {\em et~al.}, ``Giant photoinduced reflectivity modulation of nonlocal resonances in silicon metasurfaces,'' {\em Advanced Photonics}, vol.~5, no.~6, pp.~066006--066006, 2023.

\bibitem{yang2015nonlinear}
Y.~Yang, W.~Wang, A.~Boulesbaa, I.~I. Kravchenko, D.~P. Briggs, A.~Puretzky, D.~Geohegan, and J.~Valentine, ``Nonlinear fano-resonant dielectric metasurfaces,'' {\em Nano letters}, vol.~15, no.~11, pp.~7388--7393, 2015.

\bibitem{masharin2024giant}
M.~A. Masharin, T.~Oskolkova, F.~Isik, H.~Volkan~Demir, A.~K. Samusev, and S.~V. Makarov, ``Giant ultrafast all-optical modulation based on exceptional points in exciton--polariton perovskite metasurfaces,'' {\em ACS nano}, vol.~18, no.~4, pp.~3447--3455, 2024.

\bibitem{berte2024emergent}
R.~Bert{\'e}, T.~Possmayer, A.~Tittl, L.~d.~S. Menezes, and S.~A. Maier, ``Emergent resonances in a thin film tailored by optically-induced small permittivity asymmetries,'' {\em arXiv preprint arXiv:2403.05730}, 2024.

\bibitem{staude2019all}
I.~Staude, T.~Pertsch, and Y.~S. Kivshar, ``All-dielectric resonant meta-optics lightens up,'' {\em Acs Photonics}, vol.~6, no.~4, pp.~802--814, 2019.

\bibitem{sivan2020ultrafast}
Y.~Sivan and M.~Spector, ``Ultrafast dynamics of optically induced heat gratings in metals,'' {\em ACS Photonics}, vol.~7, no.~5, pp.~1271--1279, 2020.

\bibitem{Schirato2020}
A.~Schirato, M.~Maiuri, A.~Toma, S.~Fugattini, R.~Proietti~Zaccaria, P.~Laporta, P.~Nordlander, G.~Cerullo, A.~Alabastri, and G.~Della~Valle, ``Transient optical symmetry breaking for ultrafast broadband dichroism in plasmonic metasurfaces,'' {\em Nature Photonics}, vol.~14, p.~723–727, Oct. 2020.

\bibitem{Schirato2022}
A.~Schirato, G.~Crotti, R.~Proietti~Zaccaria, A.~Alabastri, and G.~Della~Valle, ``Hot carrier spatio-temporal inhomogeneities in ultrafast nanophotonics,'' {\em New Journal of Physics}, vol.~24, p.~045001, Apr. 2022.

\bibitem{Zheng2019Oct}
L.~Zheng, U.~Zywietz, A.~Evlyukhin, B.~Roth, L.~Overmeyer, and C.~Reinhardt, ``{Experimental Demonstration of Surface Plasmon Polaritons Reflection and Transmission Effects},'' {\em Sensors}, vol.~19, p.~4633, Oct. 2019.

\bibitem{dubi2019hot}
Y.~Dubi and Y.~Sivan, ````hot'' electrons in metallic nanostructures---non-thermal carriers or heating?,'' {\em Light: Science \& Applications}, vol.~8, no.~1, p.~89, 2019.

\bibitem{oguri2015dynamical}
K.~Oguri, T.~Tsunoi, K.~Kato, H.~Nakano, T.~Nishikawa, K.~Tateno, T.~Sogawa, and H.~Gotoh, ``Dynamical observation of photo-dember effect on semi-insulating gaas using femtosecond core-level photoelectron spectroscopy,'' {\em Applied Physics Express}, vol.~8, no.~2, p.~022401, 2015.

\bibitem{Schumann1991Jan}
B.~Schumann, ``{Properties of Gallium Arsenide. EMIS Datareviews Series no. 2, second edition. INSPEC, The Institute of Electric Engineering, London and New York 1990, 790 + XXIV Seiten, zahlreiche Tabellen und Literaturangaben, Sachwortverzeichnis, ISBN 0-85296-485-4},'' {\em Cryst. Res. Technol.}, vol.~26, p.~18, Jan. 1991.

\bibitem{Derrien2011}
T.~J.-Y. Derrien, T.~Sarnet, M.~Sentis, and T.~E. Itina, ``{Application of a two-temperature model for the investigation of the periodic structure formation on Si surface in femtosecond laser interaction},'' {\em J. Optoelectron. Adv. Mater.}, vol.~12, pp.~610--615, oct 2011.

\bibitem{BibEntry2001Jul}
ioffe.ru GaAs, ``{Physical properties of Gallium Arsenide (GaAs)},'' July 2001.
\newblock [Online; accessed 18. Nov. 2022].

\bibitem{Hopfel1988Apr}
R.~A. H{\ifmmode\ddot{o}\else\"{o}\fi}pfel, J.~Shah, P.~A. Wolff, and A.~C. Gossard, ``{Electron-hole scattering in GaAs quantum wells},'' {\em Phys. Rev. B}, vol.~37, pp.~6941--6954, apr 1988.

\bibitem{Sivan2020May}
Y.~Sivan and M.~Spector, ``{Ultrafast Dynamics of Optically Induced Heat Gratings in Metals},'' {\em ACS Photonics}, vol.~7, pp.~1271--1279, May 2020.

\bibitem{Margiolakis2018Dec}
A.~Margiolakis, G.~D. Tsibidis, K.~M. Dani, and G.~P. Tsironis, ``{Ultrafast dynamics and subwavelength periodic structure formation following irradiation of GaAs with femtosecond laser pulses},'' {\em Phys. Rev. B}, vol.~98, p.~224103, Dec. 2018.

\bibitem{Ruzicka2010Dec}
B.~A. Ruzicka, L.~K. Werake, H.~Samassekou, and H.~Zhao, ``{Ambipolar diffusion of photoexcited carriers in bulk GaAs},'' {\em Appl. Phys. Lett.}, vol.~97, Dec. 2010.

\bibitem{Strauss1993Jan}
U.~Strauss, W.~W. R{\ifmmode\ddot{u}\else\"{u}\fi}hle, and K.~K{\ifmmode\ddot{o}\else\"{o}\fi}hler, ``{Auger recombination in intrinsic GaAs},'' {\em Appl. Phys. Lett.}, vol.~62, pp.~55--57, Jan. 1993.

\bibitem{block2019tracking}
A.~Block, M.~Liebel, R.~Yu, M.~Spector, Y.~Sivan, F.~Garc{\'\i}a~de Abajo, and N.~F. van Hulst, ``Tracking ultrafast hot-electron diffusion in space and time by ultrafast thermomodulation microscopy,'' {\em Science advances}, vol.~5, no.~5, p.~eaav8965, 2019.

\bibitem{Sokolowski-Tinten2000Jan}
K.~Sokolowski-Tinten and D.~von~der Linde, ``{Generation of dense electron-hole plasmas in silicon},'' {\em Phys. Rev. B}, vol.~61, pp.~2643--2650, Jan. 2000.

\bibitem{Bennett1990Feb}
B.~Bennett, R.~Soref, and J.~Alamo, ``{Carrier-induced change in refractive index of InP, GaAs and InGaAsP},'' {\em Quantum Electronics, IEEE Journal of}, vol.~26, pp.~113--122, Feb. 1990.

\bibitem{Papatryfonos2021Feb}
K.~Papatryfonos, T.~Angelova, A.~Brimont, B.~Reid, S.~Guldin, P.~Smith, M.~Tang, K.~Li, A.~Seeds, H.~Liu, and D.~Selviah, ``{Refractive indices of MBE-grown Al x Ga (1{-} x ) As ternary alloys in the transparent wavelength region},'' {\em AIP Adv.}, vol.~11, p.~025327, Feb. 2021.

\bibitem{glezer1995laser}
E.~Glezer, Y.~Siegal, L.~Huang, and E.~Mazur, ``Laser-induced band-gap collapse in gaas,'' {\em Physical Review B}, vol.~51, no.~11, p.~6959, 1995.

\bibitem{singh2002ripples}
A.~P. Singh, A.~Kapoor, and K.~Tripathi, ``Ripples and grain formation in gaas surfaces exposed to ultrashort laser pulses,'' {\em Optics \& Laser Technology}, vol.~34, no.~7, pp.~533--540, 2002.

\bibitem{Liang2015Nov}
G.~Liang, Z.~Luo, K.~Liu, Y.~Wang, D.~Jianxiong, and Y.~Duan, ``{Fiber Optic Surface Plasmon Resonance-Based Biosensor Technique: Fabrication, Advancement, and Application},'' {\em Crit. Rev. Anal. Chem.}, vol.~46, p.~00, Nov. 2015.

\end{thebibliography}

\end{document}


\title[]{Excitation of surface plasmon-polaritons through optically-induced ultrafast transient gratings}

\author{Olesia Pashina}
  
  \affiliation{University of Brescia, Brescia, Italy
}%
 \affiliation{\affilITMO}

\author{Olga Sergaeva}%

\affiliation{\affilITMO 
}%

\author{Albert Seredin}%

\affiliation{\affilITMO 
}%

\author{Giulia Crotti}
\affiliation{Polytechnic University of Milan, Milan, Italy
}%
\author{Giuseppe Della Valle}
\affiliation{Polytechnic University of Milan, Milan, Italy
}%

\author{Andrey Bogdanov}
\affiliation{\affilITMO
}%

\author{Mihail Petrov}
\email{m.petrov@metalab.ifmo.ru}
\affiliation{\affilITMO
}%

\author{Costantino De Angelis}
\affiliation{University of Brescia, Brescia, Italy
}%


\begin{abstract}

Ultrafast excitation of non-equilibrium carriers   under intense pulses offer unique opportunities for controlling optical properties of semiconductor materials. In this work, we propose a scheme for ultrafast generation of surface plasmon polaritons (SPPs) via a transient metagrating formed under two interfering optical pump pulses in the semiconductor GaAs thin film. The grating can be formed due to   modulation of the refractive index associated with the non-equilibrium carriers generation. The formed temporal grating structure enables generation of SPP waves at GaAs/Ag interface  via  weak probe pulse excitation. We propose a theoretical model describing non-equilibrium carriers formation and diffusion and their contribution to permittivity  modulation via Drude and band-filling mechanisms. We predict that by tuning the parameters of the pump and probe one can reach critical coupling regime and achieve efficient generation of SPP at the times scales of 0.1-1 ps.    
\end{abstract}

\maketitle

\maketitle

\section{Theoretical model}

In this section, we take a closer look at the theoretical model describing the ultrafast interaction of a semiconductor structure with pulsed laser radiation, presented in the main text:
\begin{equation}
\begin{gathered}
\frac{\partial \mathcal{U}_{\mathrm{e}}^{\mathrm{NT}}(\mathbf{r}, t)}{\partial t}=\frac{1}{C_{\mathrm{e}}\left(N_e^{\mathrm{NT}}\right)} \nabla\left[k_{\mathrm{e}}\left(N_{\mathrm{e}}^{\mathrm{NT}}, T_{\mathrm{e}}\right) \nabla \mathcal{U}_{\mathrm{e}}^{\mathrm{NT}}\right]-\left(\Gamma_{\mathrm{e}}+\Gamma_{\mathrm{ph}}\left(N_e^{\mathrm{NT}},T_{ph}\right)\right) \mathcal{U}_{\mathrm{e}}^{\mathrm{NT}}(\mathbf{r}, t)+Q_{\mathcal{U}_{\mathrm{e}}}(\mathbf{r}, t) \\
\frac{\partial N_e^{\mathrm{NT}}(\mathbf{r}, t)}{\partial t}=Q_{{N_e}}(\mathbf{r}, t)+\nabla\left[D\left(T_e\right) \nabla N_e^{\mathrm{NT}}(\mathbf{r}, t)\right]-\Gamma_{\mathrm{e}} N_e^{\mathrm{NT}}(\mathbf{r}, t)\\
\frac{\partial N_e^{\mathrm{T}}(\mathbf{r},t)}{\partial t}=
\Gamma_e N_e^{\mathrm{NT}}(\mathbf{r},t)
+ \nabla \left[D(T_e) \nabla N_e^{\mathrm{T}}(\mathbf{r},t)\right]
-(\gamma_{nr} {N_e^{\mathrm{T}}(\mathbf{r},t)+\gamma_{r} {N_e^{\mathrm{T}}}^{2}(\mathbf{r},t)+\gamma_{Au} {N_e^{\mathrm{T}}}^{3}
(\mathbf{r},t)}\\
\frac{\partial [C_e (N_e^{\mathrm{T}}) T_e(\mathbf{r}, t)]}{\partial t} = \nabla \left[k_e(N_e^{\mathrm{T}}, T_e) \nabla T_e(\mathbf{r}, t)\right]  - G_{\mathrm{e}-\mathrm{ph}}(N_e^{\mathrm{T}})[T_e(\mathbf{r}, t)-T_{ph}(\mathbf{r}, t)] +\Gamma_{\mathrm{e}} \mathcal{U}_{\mathrm{e}}^{\mathrm{NT}}(\mathbf{r}, t)\\
\frac{\partial [C_{ph}(T_{ph})T_{ph}(\mathbf{r}, t)]}{\partial t} = \nabla[k_{ph}(T_{ph}) \nabla T_{ph}(\mathbf{r}, t)] + G_{\mathrm{e}-\mathrm{ph}}(N_e^{\mathrm{T}})[T_e(\mathbf{r}, t)-T_{ph}(\mathbf{r}, t)] +\Gamma_{\mathrm{ph}}(N_e^{\mathrm{NT}}) \mathcal{U}_{\mathrm{e}}^{\mathrm{NT}}(\mathbf{r}, t).
\end{gathered}
\label{eq_syst}
\end{equation}

where $C_e  = 3 k_B N_e$ \cite{Schumann1991Jan} is the electron heat capacity,  $k_e= (2 k_B^2 \mu_e {N_e} T_e)/{e}$ is the electron heat conductivity~\cite{Derrien2011}, 
 $k_B$ is the Boltzmann constant, $e$ is elementary charge,
 $\mathrm{\mu_{e}} = 8500~\mathrm{cm^2/(V\cdot s)}$ is the electron mobility \cite{BibEntry2001Jul},  $k_{ph}(T_{ph})$, $C_{ph}(T_{ph})$,  $\rho(T_{ph})$  are heat capacity, thermal conductivity and density for the phonon subsystem of GaAs[100] from COMSOL material library. Moreover, the rate of energy exchange between non-thermal electrons and thermal ones is denoted as $\Gamma_e=1/\tau_e$, where $\tau_e = 200~\mathrm{fs}$ is the thermalizaton rate~\cite{Hopfel1988Apr}; the rate of energy transfer between electrons and phonons~\cite{Sivan2020May} is $\Gamma_{ph}(N_e, T_{ph}) = \Gamma^{T_{ph}}=G_{\mathrm{e}-\mathrm{ph}}(N_e)/C_{ph}(T_{ph})$, where $G_{\mathrm{e}-\mathrm{ph}}(N_e) =  C_e(N_e)/\tau_{\mathrm{e-ph}}(N_e)$ is governed by electron-phonon relaxation time $\tau_{\mathrm{e-ph}} = 0.5~\mathrm{ps}\cdot[1+(N_e/2\cdot 10^{21}~\mathrm{cm^{-3})^2}]$ ~\cite{Margiolakis2018Dec}. 
 Furthermore, each differential equation considers diffusion processes, which are defined by the phonon $k_{ph}(T_{ph})$ and electron $k_e(N_e, T_e)$ heat conductivities, as well as by the coefficient of the ambipolar heat diffusion $D=2 D_{\mathrm{e}} D_{\mathrm{h}} /\left(D_{\mathrm{e}}+D_{\mathrm{h}}\right)$~\cite{Ruzicka2010Dec}, where $D_{e(h)} = k_B T_{e(h)}\mu_{e(h)}$ and $k_B$, $T$, $\mu_{e (h)}$ denote the Boltzmann constant, temperature, and mobility of electrons (holes) in the conduction (valence) band. Finally, our model also consider all possible relaxation processes in semiconductors: non-radiative decay ($\gamma_{nr}\approx 5 \cdot 10^7 \mathrm{~s}^{-1}$ \cite{Strauss1993Jan}), radiative recombination($\gamma_{r} = 1.7\cdot 10^{-10}~\mathrm{cm^3/s}$ \cite{Strauss1993Jan}), and Auger recombination ($\gamma_{Au}=7\cdot 10^{-30} \mathrm{cm^6/s}$ \cite{Strauss1993Jan}).
 
\begin{figure*}[h!]
    \centering
    \includegraphics[width=1\columnwidth]{Figs/fig_s1_2iter.pdf}
    \caption{a) During electromagnetic excitation of a semiconductor, the energy of incident photons is absorbed by the electrons within the material, causing interband transitions from the valence band to the conduction band.  b) Diagram illustrating the physical processes taking place under the ultrafast irradiation of a semiconductor structure. c) A self-consistent approach involves solving a system of differential equations describing the evolution of key characteristics while considering Maxwell's equations. d) Band diagram illustrating the processes of thermalization and absorption.}  
     \label{fig_supl_scheme}
\end{figure*}

This self-consistent system of differential equations chracterizes spatial dynamics of the energy of non-thermalized electrons $\mathcal{U}_{\mathrm{e}}^{ \mathrm{NT}}( \mathbf{r}, t)$, their concentration ${N}_{\mathrm{ e}}^{\mathrm{NT}}(\mathbf{r}, t)$, the concentration of thermalized ones  ${N}_{\mathrm{ e}}^{\mathrm{T}}(\mathbf{r}, t)$. Moreover, the last two expressions determine the evolution of temperatures  ${T}_{\mathrm{ e}}(\mathbf{r}, t)$ and ${T}_{\mathrm{ph}}(\mathbf{r}, t)$, which describes the electron and phonon systems, introducing the dynamics of their respective energies.

Now let's analyze the primary physical effects and optical processes underlying the mentioned coupled differential equations (FIG.~\ref{fig_supl_scheme}b-c)), and also determine the corresponding characteristic times of these processes.

\subsection{Generation of non-thermalized electrons}
First of all, when electromagnetic radiation interacts with a semiconductor, incident photons transfer their energy to material's electrons, prompting them to transition from the valence band to the conduction one. This process results in the creation of electron-hole pairs and can be viewed as the generation of non-thermalized electrons within the conduction band. Thus, these electrons exhibit an increase in both their energy $\mathcal{U}_{\mathrm{e}}^{\mathrm{NT}}$ and concentration ${N}_{\mathrm{e}}^{\mathrm{NT}}$. Consequently, the energy source of non-thermalized electrons $Q_{\mathcal{U}_{\mathrm{e}}}(\mathbf{r})$ can be accurately determined by calculating the absorbed power for a specific coordinate $\mathbf{r}$:
\begin{equation}
P_{\mathrm{abs}}(\mathbf{r})=\frac{1}{2} \mathrm{Re}\left[\mathbf{J}^*(\mathbf{r}) \cdot \mathbf{E(r)} \right]=\frac{1}{2} \mathrm{Re}\left[i\omega(\varepsilon-1)^*\varepsilon_0\mathbf{E}(\mathbf{r})^*\cdot\mathbf{E} (\mathbf{r})\right]= \frac{1}{2}\omega  \mathrm{Im}[\varepsilon]\varepsilon_0|\mathbf{E}(\mathbf{r})|^2,
\label{P_abs_r}
\end{equation}
where $\varepsilon_0$ is the permittivity of vacuum, $\omega$ is the frequency of electromagnetic radiation corresponding to the dielectric permittivity $\varepsilon$,
$
\mathbf{E}(\mathbf{r}, t)= \mathrm{Re}\left[\mathbf{E}(\mathbf{r}) e^{-\mathrm{i} \omega t}\right]$ is the electric field, is $\mathbf{J}(\mathbf{r}, t)=\mathrm{Re}\left[\mathbf{J}(\mathbf{r}) e^{-\mathrm{i} \omega t}\right]
$ is the current. Here we utilize that the current is given by $\mathbf{J}=\partial\mathbf{P}/\partial t=-i\omega \mathbf{P} = -i\omega(\varepsilon-1)\varepsilon_0 \mathbf{E}$, according to the connection between displacement field $\mathbf{D}$ and polarization $\mathbf{P}$ in the form $\mathbf{D}=\varepsilon\varepsilon_0\mathbf{E} = \varepsilon_0\mathbf{E} +\mathbf{P}$.

Since in our model we irradiate the GaAs film with a plane-wave Gaussian pulses, in order to define time-dependent sources, one should multiply equation (\ref{P_abs_r}) by the Gaussian pulse shape in terms of FWHM $ \mathrm{GP}(t) = \mathrm{exp} (-4 \cdot \mathrm{ln}2\cdot(t-t_0)^2/\tau_\mathrm{p}^2)$, where $t_0$  represents the peak time of the pulse, $\tau_\mathrm{p}$ relates to the full width at half-maximum (FWHM) pulse duration. In this work, we focus on the specific peak time of the pulse $t_0 = 0$ ps  and the full width at half-maximum (FWHM) pulse duration $\tau_\mathrm{p} = 100$ fs.
Finally, we obtain the following equation for the energy source at each point within the semiconductor structure  \begin{equation}
Q_{\mathcal{U}_\mathrm{e}}(\mathbf{r},t) = P_\mathrm{abs}(\mathbf{r})\cdot \mathrm{GP}(t) = \frac{1}{2}\omega\mathrm{Im}[\varepsilon(\mathbf{r}, t)]\varepsilon_0|\mathbf{E}(\mathbf{r})|^2\cdot \mathrm{GP}(t).
\label{eq_source}
\end{equation}
In this expression, we consider the dielectric permittivity to be dependent not only on spatial location $\mathbf{r}$, but also on the specific time moment $t$. This assumption aligns with the dynamic nature of dielectric permittivity, which evolves over time. A detailed explanation of this evolution will be provided in the subsequent chapter (See Supplemental Information, Sec. 2).

At the same time, as long as each photon results in the generation of one non-thermalized electron (namely an electron-hole pair), to determine the source of concentration, one can divide the energy source by the electromagnetic energy of one photon and obtain
$Q_{N_e}(\mathbf{r},t) = Q_{\mathcal{U}_{\mathrm{e}}}(\mathbf{r},t)/h\omega$.
In this study, we employ the pump-probe technique to investigate the all-optical formation of a diffraction grating and the consequent excitation of the surface plasmon polariton. Notably, it is the pump excitation, characterized by high power compared to the probe, that drives the evolution of the main characteristics specified in the differential equations~[\ref{eq_syst}], thus in the sources we utilize the pump frequency $\omega$ corresponding to a specific wavelength $\lambda_\mathrm{pump} = 500$ nm.

\subsection{Thermalization}

Returning to the analysis of optically induced processes, after the generation of non-thermalized carriers, their energy begins to dissipate due to the exchange of energy of non-thermalized carriers with thermal electrons and phonons (FIG.~\ref{fig_supl_scheme}d)). The rate of energy exchange between non-thermal and thermal electrons is denoted as $\Gamma_e=1/\tau_e$ and represents the thermalization process of the electron distribution as a whole. Previous research has demonstrated that this energy relaxation time, resulting from electron-electron scattering, ranges from several hundred femtoseconds (200-500 fs) depending on the electron temperature (300-500 K)~\cite{Hopfel1988Apr}. So, in our study we assume that $\tau_e = 200~\mathrm{fs}$.  

The thermalization process leads to a reduction in the concentration of non-thermalized electrons and a simultaneous increase in the concentration of thermalized electrons, denoted by the term $\Gamma_e N_e^{\mathrm{NT}}(\mathbf{r},t)$. Moreover, this effect also results in an elevation of energy for thermalized electrons by $\Gamma_{\mathrm{e}} \mathcal{U}_{\mathrm{e}}^{\mathrm{NT}}$. It is important to highlight here that the energies of the phonon and thermalized electron subsystems are typically governed by the product of their respective heat capacities $C$ and temperatures $T$. Therefore, we introduce the energies  $C_e (N_e^{\mathrm{T}}) T_e(\mathbf{r}, t)$ and $C_{ph}(T_{ph})T_{ph}(\mathbf{r}, t)$ for the phonon and thermalized electron subsystems and consider their dynamics in the last two differential equations of the system. Here, $C_e(N_e)  = 3 k_B N_e$ \cite{Schumann1991Jan} is the electron heat capacity, while $C_{ph}(T_{ph})$ denotes the lattice heat capacity of GaAs[100] from COMSOL material library.  For room temperature $T_{ph}=300$ K, one can utilize the lattice heat capacity $C_{ph} = 330~~\mathrm{J/(kg\cdot K)}\cdot \rho$, where $\mathrm{\rho = 0.532~\mathrm{kg/cm^3}}$ - is the density~\cite{BibEntry2001Jul}. Subsequently, the evolution of temperature and concentration within the thermal electron subsystem due to thermalization can be described by the terms $\Gamma_e \mathcal{U}_e^{\mathrm{NT}}(\mathbf{r},t)$ and $\Gamma_e N_e^{\mathrm{NT}}( \mathbf{r},t)$, respectively.

\subsection{Electron-phonon scattering}
Furthermore, both thermalized and non-thermalized electrons not only exchange energy with each other during the process of thermalization, but also transfer energy to the phonon subsystem through electron-phonon scattering. This physical process can be described by the energy exchange rate  $G_{\mathrm{e}-\mathrm{ph}}(N_e) =  C_e(N_e)/\tau_{\mathrm{e-ph}}(N_e)$, where $\tau_{\mathrm{e-ph}}(N_e) = 0.5~\mathrm{ps}\cdot[1+(N_e/2\cdot 10^{21}~\mathrm{cm^{-3})^2}]$ is the electron-phonon relaxation time~\cite{Margiolakis2018Dec}. This rate indicates distinct time scales for the temperature evolution of the thermalized electron and phonon systems, which can be approximated as $\Gamma^{T_{e}}(N_e)=G_{\mathrm{e}-\mathrm{ph}(N_e)}/C_{e}(N_e)$ and $\Gamma^{T_{ph}}(N_e, T_{ph})=G_{\mathrm{e}-\mathrm{ph}}(N_e)/C_{ph}(T_{ph})$, respectively. 

Moreover, due to the identical physical nature of electron-phonon scattering process for thermalized and non-thermalized electrons, the characteristic time scale of energy transfer between non-thermalized electrons and phonons can be defined similarly to one for thermalized electrons. It can be represented as $\Gamma_{ph}(N_e, T_{ph}) = \Gamma^{T_{ph}}(N_e, T_{ph})=G_{\mathrm{e}-\mathrm{ ph}}(N_e)/C_{ph}(T_{ph})$, where in this case $N_e$ relates to the concentration of non-thermalized electrons. Therefore, this brings us to the terms $\Gamma_{\mathrm{ph}}(N_e^{\mathrm{NT}}, T_{ph}) \mathcal{U}_{\mathrm{e}}^{\mathrm{NT}}( \mathbf{r}, t)$ for energy transfer between the lattice and non-thermalized electrons and $G_{\mathrm{e}-\mathrm{ph}}(N_e^{\mathrm{T}})[T_e(\mathbf{ r}, t)-T_{ph}(\mathbf{r}, t)]$ for interaction between phonons and thermalized electrons~\cite{Sivan2020May}. 

For $T_{ph} = 300$ K and different values of electron-hole plasma concentration $N_e$, $\Gamma^{T_{ph}}(N_e)$ or $\Gamma_{ph}(N_e)$  range from $7.5\cdot 10^{-5}~\mathrm{s^{-1}}$ [for initial intrinsic electron concentration in the conduction band $N_e^0 = 2.1\cdot10^{6}~\mathrm{cm^{-3}}$] to $3.5\cdot 10^{9}~(4.7\cdot 10^{10})~\mathrm{s^{-1}}$ (several hundred (tens) picoseconds) [for high-density electron-hole plasma $N_{e} = 10^{20} ~(2\cdot 10^{21})~\mathrm{cm^{-3}}$]. On the other hand, $\Gamma^{T_{e}}(N_e)$ also slightly depends on the electron concentration in the conduction band and takes the value $2\cdot 10^{12}~\mathrm{s^{-1}}$ [for the initial intrinsic electron concentration in the conduction band $N_e^ 0 = 2.1\cdot10^{6}~\mathrm{cm^{-3}}$] to $2\cdot 10^{12} ~(1 \cdot 10^{12})~\mathrm{s^{-1}}$ (several hundred femtoseconds (picosecond)) [for high-density electron-hole plasma $N_{e} = 10^{20} ~(2\cdot 10^{21})~\mathrm{cm^{-3}}$]. It can be noted that, as a result, for any concentration of electron-hole plasma $\Gamma^{T_{e}}\gg\Gamma^{T_{ph}}$. 
Moreover, it can be observed that $\Gamma_{e}~\textless~ \Gamma^{T_{e}}$ and $\Gamma_{e} \ll \Gamma^{T_{ph}}, \Gamma_{{ph}}$, indicating that the thermalization processes are significantly faster compared to other evolutions resulting from electron-phonon scattering.
\subsection{Relaxation}
In semiconductor physics, another crucial physical phenomenon is the relaxation process of electrons, which entails the recombination of an electron in the conduction band with a hole (an electron vacancy), resulting in the annihilation of both charge carriers. Although relaxation effects for non-thermalized electrons are typically disregarded due to rapid thermalization, these processes become significant when analyzing temporal characteristics of thermalized electrons. Therefore, relaxation processes in semiconductors can be categorized into three types: non-radiative decay, radiative recombination, and Auger recombination.

Non-radiative decay involves the transfer of energy without the emission of light. Such trap-assisted processes include Shockley-Read-Hall recombination and surface recombination. Shockley-Reed-Hall recombination corresponds to lattice defects or unintentional impurity atoms. In this effect an electron or hole is first trapped by a defect and then emitted into the valence/conduction band, resulting in a decrease in the concentration of charge carriers in the corresponding band. Recombination at surfaces is almost identical to Shockley-Hall-Reed recombination, but in this case recombination occurs specifically at surfaces, which typically contain a large number of recombination centers due to the abrupt termination of the semiconductor crystal. Additionally, surfaces and interfaces are more prone to contain impurities because of exposure during the fabrication process. Finally, in our model we utilize the volume nonradiative trap-assisted recombination coefficient $\gamma_{nr}\approx 5 \cdot 10^7 \mathrm{~s}^{-1}$ \cite{Strauss1993Jan}. Moreover, at the interface between air and semiconductor material we determine the surface non-radiative relaxation coefficient $\gamma_{surf}\approx 5 \cdot 10^6\mathrm{~cm/s}$ \cite{Ito1994Jan}.

Radiative recombination is the recombination of electron-hole pairs in a semiconductor, accompanied by the emission of photon with energy close to the band gap. In the case of GaAs, the radiative recombination can be characterized by a rate constant of $\gamma_{r} = 1.7\cdot 10^{-10}~\mathrm{cm^3/s}$ \cite{Strauss1993Jan}.

Moreover, the time evolution of the concentration takes into account Auger recombination, a non-radiative process involving three charge carriers. In this process, excess energy from the recombination of electron-hole pairs is transferred to other electrons or holes, causing them to move to higher energy states in the same zone instead of emitting photons. In many works studying GaAs, the Auger coefficient varies around the values of $\gamma_{Au}=10^{-31 \pm 1}~\mathrm{cm^6/s}$ \cite{Steiauf2014Aug}. Therefore, in our model we use the Auger coefficient, that falls within the experimentally determined range, specifically $\gamma_{Au}=7\cdot 10^{-30} \mathrm{cm^6/s}$ \cite{Strauss1993Jan}.

One evident implication of the discussed recombination coefficients in GaAs material is to define the predominant recombination mechanisms across varying electron concentration ranges. It is notable that nonradiative recombination prevails at concentrations below $N_e\ll10^{18}~\mathrm{cm^{-3}}$. As the concentration of injected electrons within the GaAs structure increases, reaching values around $N_e^{max}\approx 5\cdot 10^{18} ~\mathrm{cm ^{-3}}$, the radiative term begins to exceed other relaxation mechanisms. Subsequently, as the concentration reaches approximately $N_e\approx 10^{19-20}~\mathrm{cm^{-3}}$, the rate of concentration decrease  begins to be determined mainly by cubic Auger recombination. This analysis highlights the importance of considering all recombination processes to accurately control the behavior of conduction electrons in GaAs under ultrafast excitation, covering all three concentration ranges.

\subsection{Diffusion}

At the same time, it is important to consider carrier transport resulting from diffusion. 
Indeed, both themalized and non-thermalized carriers
diffuse from high-density to low-density regions, leading to an expansion of the carrier density profile. Due to the Coulomb interaction, the excited electrons in the conduction band and positively charged holes in the valence band diffuse together as pairs. The coefficient for this ambipolar diffusion can be calculated as 
$D (T_e)=2 D_{\mathrm{e}} D_{\mathrm{h}} /\left(D_{\mathrm{e}}+D_{\mathrm{h}}\right)$~\cite{Ruzicka2010Dec}, where $D_e$ and $D_h$ represent the diffusion coefficients of electrons and holes respectively. These coefficients can be determined using the Einstein relation, which states that $D_{e(h)} = k_B T_{e(h)} \mu_{e(h)}$, where $k_B$, $T$, and $\mu_{e(h)}$ denote the Boltzmann constant, temperature, and electron (hole) mobility in the conduction (valence) band. The ambipolar diffusion coefficient of photoexcited carriers in intrinsic GaAs at room temperature is $D = 20~ \mathrm{cm^2/s}$~\cite{Ruzicka2010Dec}. Finally, the diffusion of electron concentration can be characterized by the concentration diffusion time $\tau_\mathrm{diff}^{N_e} = \frac{d^2}{4\pi^2D}$, where $d$ represents the characteristic length scale associated with the initial concentration nonuniformity, and $D$ is the diffusion coefficient. Assuming $d\approx 1~\mathrm{\mu m}$ (since the period of our periodic grating is $a=1100$ nm), the concentration diffusion parallel to the GaAs film can extend to tens of picoseconds for room temperature $T_e = 300$ K. 
However, with the increase in electron temperature, this process accelerates, corresponding to $\tau_\mathrm{diff}^{N_e} = 390$ fs for $T_e = 10^4$~K.

Moreover, similar to concentration diffusion, heat diffusion should be considered for both the thermalized electron and phonon subsystems. Assuming the non-uniformity of the induced field within the film, the electron heat diffusion time can be approximated as $\tau_\mathrm{diff}^{T_e} = \frac{d^2}{4\pi^2D_e}$, where $D_e (T_e) = \frac{k_e}{C_e} = (2k_B\mu_eT_e)/(3e)$ denotes the heat diffusion coefficient for electrons. This coefficient can reach values of approximately $D_e \approx 147~\mathrm{cm^2/s}$ at room temperature $T_e= 300$ K, and it increases with an elevation in electron temperature. Importantly, when the illumination inhomogeneities are sufficiently strong (i.e., $d\approx 1~\mathrm{\mu m}$), the diffusion of electron temperature parallel to the GaAs film can occur within a few picoseconds, which is an order of magnitude faster than the concentration diffusion of electrons. Furthermore, as $T_e$ grows, the thermal diffusion coefficient $D_e$ also increases, resulting in even smaller time scales for diffusion of electron temperature (about $\tau_\mathrm{diff}^{T_e}=52~\mathrm{fs}$ for $T_e = 10^4$~K). 

Additionally, we can introduce the phonon heat diffusion time $\tau_\mathrm{diff}^{T_{ph}} = \frac{d^2}{4\pi^2D_{ph}}$, where $D_{ph} \equiv \frac{k_{ph}}{C_{ph}}=0.31 ~\mathrm{cm^2/s} $ close to room temperature, since $C_{ph} = 330~~\mathrm{J/(kg\cdot K)}\cdot \rho$, $k_{ph} = 0.55~\mathrm{W/(cm\cdot K)}$, and $\mathrm{\rho = 0.532~\mathrm{kg/cm^3}}$ - are the lattice heat capacity, thermal conductivity, and density, respectively for $T_{ph}\approx300~\mathrm{K}$~\cite{BibEntry2001Jul}. Thus, for non-uniformity with $d\approx 1~\mathrm{\mu m}$, this phenomenon is observed on a time scale of approximately one hundred nanoseconds. Comparing the heat diffusion for the electron and phonon subsystems, one can notice that $\tau_\mathrm{diff}^{T_{e}}\ll\tau_\mathrm{diff}^{T_{ph}}$.  

Furthermore, we consider the heat diffusion of non-thermalized electrons by assuming that the diffusion coefficient for both non-thermalized and thermalized electrons is identical, as demonstrated in previous studies on metals~\cite{Block2019May}.  Consequently, we employ the expression $D_{U_e^{\mathrm{NT}}}=k_e(N_e^{NT})/C_e(N_e^{NT})$ to characterize this phenomenon.

\subsection{Characteristic time scales}
Table~\ref{t} illustrates the characteristic time scales of various optically-induced processes underlying the developed theoretical model for the ultrafast interaction of GaAs material with pulsed electromagnetic radiation.
\begin{center}
\begin{table}[ht]

\centering
\begin{tabular}{| m{5.5cm} | m{4cm}| m{2.35cm}| } 
 \hline
\textbf{Effect}  & \textbf{Characteristic time scale} &  \textbf{Value}\\
 \hline
  \centering
Thermalization of non-thermal electrons $\mathcal{U}_{\mathrm{e}}^{\mathrm{NT}}$ & \centering $\tau_{e}=1/\Gamma_{e}(T_e)$  & $\approx 200$ fs \\
  \hline
    \centering
   Evolution of electron temperature $T_e$ due to electron-phonon scattering  & \centering $1/\Gamma^{T_{e}}(N_e)$  & $\approx 500$ fs \\
  \hline
    \rule{0cm}{0.45cm}
   Diffusion of electron temperature $T_e$ & \centering $\tau_\mathrm{diff}^{T_e}(d, T_e)$  & $\approx2$ ps - 52 fs \\ 
  \hline
      \rule{0cm}{0.45cm}
  Diffusion of electron concentration $N_e$  & \centering $\tau_\mathrm{diff}^{N_e}(d, T_e)$  &    $\approx13$ ps - 390 fs \\
 \hline
   \centering
Evolution of phonon temperature $T_{ph}$ due to electron-phonon scattering  & \centering $1/\Gamma^{T_{ph}}(N_e)$  & $\approx 0.3$ ns \\
  \hline
  \centering
  Energy transfer between non-thermal electrons and phonons  & \centering $1/\Gamma_{ph}(N_e)$  & $\approx 0.3$ ns\\
  \hline
 \rule{0cm}{0.45cm}
   Diffusion of phonon temperature $T_{ph}$ & \centering $\tau_\mathrm{diff}^{T_{ph}}(d)$  & $\approx 80$ ns\\ 
  \hline

    \end{tabular}
    \label{t}
    \caption{Sorted in ascending order characteristic time scales of various ultrafast processes underlying the two-temperature model of the GaAs material. These values are defined for room electron temperatures $T_e=300-10^4~\mathrm{K}$ and high-density electron-hole plasma $N_{e} = 10^{20}~\mathrm {cm^{- 3}}$ taking into account the characteristic nonuniformity length scale $d \approx 1~{\mu \mathrm{m}}$.}
\end{table}
\end{center}

\section{Dielectric permittivity modulation}
Returning to the system of differential equations, we can notice that the sources of concentrations and energy are determined by the dielectric constant of the structure. Indeed, on the one hand, they are proportional to the imaginary part of the dielectric constant, and on the other hand, the field induced inside the structure also depends on the latter’s dielectric constant. However, it turns out that over time the dielectric constant of the semiconductor itself evolves due to the optically induced generation of nonequilibrium carriers in the conduction band. The dynamics of optical properties are determined by two main processes - the generation of free carriers, which is described by the Drude model, as well as taking into account the band filling effect. Thus, the dielectric constant changing with time introduces a high degree of self-consistency in our model.

Returning to the system of differential equations~\ref{eq_syst}, it becomes apparent that the concentration  $Q_{N_\mathrm{e}}(\mathbf{r},t)$ and energy sources $Q_{\mathcal{U}_\mathrm{e}}(\mathbf{r},t)$ are influenced by the dielectric constant of the structure  $\varepsilon(\mathbf{r},t)$ (See Eq.~\ref{eq_source}). Indeed, on the one hand, they are proportional to the imaginary part of the dielectric permettivity $\mathrm{Im}~\varepsilon(\mathbf{r},t)$, and on the other hand, the field induced inside the structure $\mathbf{E}(\mathbf{r},t)$ also strongly depends on the dielectric constant. 

Notably, under electromagnetic irradiation this dielectric permittivity undergoes changes over time due to the optically induced generation of nonequilibrium carriers in the conduction band of the semiconductor. The evolution of optical properties is governed by two primary mechanisms: the generation of free carriers, as outlined in the Drude model, and the consideration of the band filling effect. As a result, the temporal variation of the dielectric constant introduces a higher level of self-consistency to our model. Indeed, the initial optical-induced modulation of the electron concentration $N_e$ leads to a change in the complex dielectric constant, which, in turn, affects the subsequent electromagnetic absorption and the values of the sources that determine the concentration $N_e$ and other key characteristics of the semiconductor.

Let's delve into a detailed examination of both the Drude and band filling effects and analyze their impact on dielectric modulation.

\subsection{Band-filling effect}
The band-filling effect is a phenomenon that arises as a result of optically induced interband transitions of electrons from the valence band to the conduction band and consists of filling low-energy states inside the conduction band. As mentioned earlier, when a material is irradiated with electromagnetic radiation at a frequency above the bandgap, the electron concentration in the conduction band increases due to absorption, which leads to the gradual filling of the lowest energy states in the conduction band. The latter causes a decrease in the probability of subsequent optically induced transitions of electrons from the valence band to the conduction band in accordance with the Pauli principle, which prohibits transitions between occupied states. Consequently, the absorption coefficient of photons with energies above the band gap gradually decreases. Therefore, the interband absorption coefficient, which considers the band filling effect, can be determined using the following equation:

\begin{equation} 
\alpha_{\mathrm{ib}} = \alpha_{\mathrm{ib0}} + \Delta\alpha_{bf}.
\end{equation}

Here, $\alpha_{\mathrm{ib0}}$ represents the initial interband absorption coefficient, and $\Delta\alpha_{bf}$ is a negative correction term accounting for the impact of the filling effect. 

The initial interband absorption coefficient included transitions from heavy and light hole valence bands can be calculated as  ~\cite{Bennett1990Feb} 
\begin{equation}
\alpha_{\mathrm{ib0}}\left(N_e, E,T\right) = \alpha_{\mathrm{ib0}}^{hh}\left(N_e, E,T\right)+\alpha_{\mathrm{ib0}}^{lh}\left(N_e, E,T\right),
\end{equation}
\begin{equation}
\begin{aligned}
\alpha_{\mathrm{ib0}}^{hh}\left(N_e, E,T\right) \mathrm{[cm^{-1}]}= 
\frac{C_{h h}}{E \mathrm{[eV]}} \sqrt{E\mathrm{[eV]}-E_g\mathrm{[eV]}}{\sqrt{\frac{h}{2 \pi e}}},\\ \alpha_{\mathrm{ib0}}^{lh}\left(N_e, E,T\right) \mathrm{[cm^{-1}]}= \frac{C_{l h}}{E\mathrm{[eV]}} \sqrt{E\mathrm{[eV]}-E_g\mathrm{[eV]}}{\sqrt{\frac{h}{2 \pi e}}},
\end{aligned}
\end{equation}
where $E$ is the photon energy, $E_g$ is the energy gap, $h$ is Plank constant, $e$ is elementary charge. The coefficient $\sqrt {\frac{h}{2 \pi e}}$ is not mentioned in~\cite{Bennett1990Feb}, but added for consistency units of measurement and finally the mass constants for heavy and light holes are as follows
\begin{equation}
C_{h h}=C\left(\frac{\mu_{e h h}^{3 / 2}}{\mu_{e h h}^{3 / 2}+\mu_{e l h}^{3 / 2}}\right),\quad 
C_{lh}=C\left(\frac{\mu_{e l h}^{3 / 2}}{\mu_{e h h}^{3 / 2}+\mu_{e l h}^{3 / 2}}\right).
\end{equation}
Here constant $C=2.3 \cdot 10^{12} ~\mathrm{cm^{-1} s^{-1/2}}$ for GaAs~\cite{Bennett1990Feb} and the reduced effective masses of the electron-hole pairs are defined as
\begin{equation}
\mu_{e h h}=\left(\frac{1}{m_e}+\frac{1}{m_{h h}}\right)^{-1},\quad
\mu_{e l h}=\left(\frac{1}{m_e}+\frac{1}{m_{l h}}\right)^{-1},
\end{equation}
where $m_e$ - electron mass, $m_{lh}$ - mass of light holes. Here and below, we will take $T$ to be $T_{ph}$ - the phonon temperature.

Finally, the change in the interband absorption coefficient due to the band filling effect can be described through the probability densities of state occupations as follows~\cite{Bennett1990Feb} :
\begin{equation}
\begin{aligned}
\Delta \alpha_{bf}\left(N_e, E,T\right) = 
&\alpha_{\mathrm{ib0}}^{hh}\left(N, E,T\right)\left[f_v(E_{ah})-f_c(E_{bh})-1\right]\\
+&\alpha_{\mathrm{ib0}}^{hh}\left(N, E,T\right) \left[f_v(E_{al})-f_c(E_{bl})-1\right],
\end{aligned}
\end{equation}
where the indices $h$ and $l$ represent the heavy-hole and light-hole valence bands, respectively. One can note that this modification in interband absorption will depend on the electron concentration $N_e$ and phonon temperature $T_{ph}$, as well as the initial coefficient.

The probability of a conduction band state of energy $E_b$ being occupied by an electron is denoted as $f_c(E_b)$, and the probability of a valence band state of energy $E_a$ being occupied by an electron corresponds to $f_v(E_a)$. These probabilities are determined by the Fermi-Dirac distribution functions $f_{c}(E)=[1+e^{(E-E_{F_c})/(k_B T)}]^{-1}$ and $f_{v}(E)=[1+e^{(E-E_{F_v})/(k_B T)}]^{-1}$.

By defining the energy of one photon of incident radiation as $E=\hbar\omega$, we can obtain that the energies of potential states in the conduction band and valence band, which can be involved in possible optical induced transitions, are as follows:
\begin{equation}
\begin{aligned}
E_{ah, al}(E)&= (E_g-E)\left( \frac{m_e}{m_e+m_{hh,lh}}\right)-E_g,\\
E_{bh, bl}(E)&= (E-E_g)\left( \frac{m_{hh,lh}}{m_e+m_{hh,lh}}\right).
\end{aligned}
\end{equation}

This follows from the laws of conservation of energy and momentum~\cite{Bennett1990Feb}.
At the same time, the carrier-dependent quasi-Fermi levels are estimated by the Nilsson approximation~\cite{Nilsson1978Oct}:
\begin{equation}
\begin{aligned}
E_{F_c}(N_e,T)=&\left\{\ln \left(\frac{N_e}{N_c}\right)+\frac{N_e}{N_c}\left[64+0.05524 \frac{N_e}{N_c}\cdot \left(64+\sqrt{\frac{N_e}{N_c}}\right)\right]^{-(1 / 4)}\right\} k_B T, \\
E_{F_v}(N_e,T)=-&\left\{\ln \left(\frac{N_e}{N_v}\right)+\frac{N_e}{N_v}\left[64+0.05524 \frac{N_e}{N_v}\cdot \left(64+\sqrt{\frac{N_e}{N_v}}\right)\right]^{-(1 / 4)}\right\} k_B T -E_g
\end{aligned},
\end{equation}
where the zero of energy is defined to be at the conduction
band minimum. Moreover, here the effective densities of states in the conduction band and valence band are defined as $N_c(T)=2\left(\frac{m_e k_B T}{2 \pi \hbar^2}\right)^{3 / 2} $ and $
N_v(T)=2\left(\frac{m_{dh} k_B T}{2 \pi \hbar^2}\right)^{3 / 2}$ respectively, $k_B$ - Boltzmann constant, $\hbar$ - reduced Plank constant and $m_{dh} = (m_{ hh }^{3/2}+m_{lh}^{3/2})^{2/3}$ is the effective mass of the hole density of states~\cite{Bennett1990Feb}. 

Finally, the change in the extinction coefficient due to band filling is determined from the relationship between the absorption and extinction coefficients $\alpha = 4\pi k/\lambda$:
\begin{equation}
\Delta k_{bf}(N_e,E,T) = \frac{ \lambda\mathrm{[cm]}\cdot \Delta \alpha_{bf}\left(N_e, E,T\right)\mathrm{[cm^{-1}]}}{4\pi}.
\label{eq_k}
\end{equation}
On the other hand, the change in the refractive index can be calculated by integrating the resulting change in the extinction coefficient based on the Kramers-Kronig formulas:
\begin{equation}
\Delta n_{bf}(N_e, E\mathrm{[eV]}, T)=
\frac{h}{2 \pi^2 e}
P \int_0^{\infty} \frac{\Delta \alpha_{bf}\left(N_e, E^{\prime},T\right)\mathrm{[cm^{-1}]}}{E^{\prime 2}-E^2} d E^{\prime}
\label{eq_n}
\end{equation}
Here we use a slightly different coefficient from \cite{Bennett1990Feb} to match the units of measurement and get results similar to the article.

Finally, in the equations governing the modulation of the complex refractive index $\Delta n_{bf} +i\Delta k_{bf}$, we assume that $E = \hbar\omega$ represents the incident photon energy, $T$ signifies the spatiotemporal phonon temperature of structure $T_{ph}(\mathbf{r},t)$, and $N_e$ denotes the spatiotemporal concentration of thermalized electrons $N_e^T(\mathbf{r},t)$, due to the rapid nature of the thermalization process.

Based on the given changes in the refractive indices and extinction coefficient of GaAs material, we obtain the following complex dielectric constant, taking into account the effect of band filling:

\begin{equation}
\varepsilon_{bf}(\mathbf{r},t) = [(n_0+\Delta n_{bf}(\mathbf{r},t)) +i(k_0+\Delta k_{bf}(\mathbf{r},t))]^2=\varepsilon^0+\mathrm{Re} (\Delta\varepsilon_{bf}(\mathbf{r},t)+i\mathrm{Im} (\Delta\varepsilon_{bf}( \mathbf{r},t),
\end{equation}
where $n_0+ik_0$ is the initial complex refractive index and $\varepsilon^0$ is the corresponding initial complex dielectric permittivity of GaAs~\cite{Papatryfonos2021Feb}.

\subsection{Free carriers}
As previously mentioned, when exposed to ultrafast laser radiation, semiconductors  experience the generation of non-equilibrium electrons in the conduction band. These excited electrons can be regarded as charge carriers that freely move within the semiconductor material. Therefore, the Drude model, typically used to explain the dielectric constant in metals, can also be extended to these free electrons in semiconductors.

Consequently, we can observe an additional contribution to the dielectric constant according to the Drude model, resulting from free electron generation, as follows ~\cite{Sokolowski-Tinten2000Jan}
\begin{equation}
\Delta\varepsilon_{Dr}(\mathbf{r},t)=\mathrm{Re} (\Delta\varepsilon_{Dr}(\mathbf{r},t)+i\mathrm{Im} (\Delta\varepsilon_{Dr}( \mathbf{r},t), 
\end{equation}
where the Drude modulation of the real and imaginary parts for the permittivity
\begin{equation}
\begin{aligned}
\mathrm{Re} (\Delta\varepsilon_{Dr}(N_e,\omega))
=-\frac{{\omega_{p}}^2}{\omega^2+1/\tau_{\mathrm{e}-\mathrm{e}}^{2}},\\
\mathrm{Im} (\Delta\varepsilon_{Dr}(N_e,\omega))
=\frac{{\omega_{p}}^2
\tau_{\mathrm{e}-\mathrm{e}}}{\omega(1+\omega^2\tau_{\mathrm{e}-\mathrm{e}}^2)},
\end{aligned}
\end{equation}
with the plasma frequency 
\begin{equation} 
\omega_p(N_e)=\sqrt{\frac{\delta N_ee^2}{\varepsilon_{0} m_{\mathrm{eff}}}}.
\end{equation}
 Here $\omega = 2 \pi c / \lambda$ - is the frequency of incident electromagnetic radiation, $\tau_{e-e}\approx 80~\mathrm{fs}$~\cite{Hopfel1988Apr} is the characteristic  electron-electron scattering time, $\delta N_e=  N_e-N_e^0$ - the free carriers' concentration increase, 
$\mathrm{m_{eff} = 0.066m_e}$~\cite{BibEntry2001Jul} is the effective mass of electrons for GaAs,
$\varepsilon_0$ is the permittivity of vacuum, and
$e$ is the elementary charge. Finally, it is important to note that $N_e$ represents the spatiotemporal concentration of thermalized electrons $N_e^T(\mathbf{r},t)$ and defines the dependence of dieectric permettivity modulation $\Delta\varepsilon_{Dr}(\mathbf{r},t)$ on the spatial position within the structure $\mathbf{r}$ and time $t$.

\subsection{Impacts of contributions to self-consistency}
The final spatiotemporal dielectric constant of a semiconductor material under the influence of ultrafast laser radiation, which takes into account the effect of band filling and modulation associated with the generation of free electrons according to the Drude model, takes the following form
\begin{equation}
\varepsilon_\mathrm{fin}(\mathbf{r},t) = \varepsilon^0+\Delta\varepsilon_{bf} (\mathbf{r},t)+ \Delta\varepsilon_{Dr}(\mathbf{r},t),
\end{equation}
where $\Delta\varepsilon_{bf} (\mathbf{r},t)$ and $\Delta\varepsilon_{Dr} (\mathbf{r},t)$ represent the complex modulations of dielectric permittivity due to the band filling effect and Drude model, respectively.

The FIG.~\ref{fig_supl_bf_dr} illustrates the modulation in the real part $\mathrm{Re} (\Delta\varepsilon_{Dr})$ and imaginary parts $\mathrm{Im} (\Delta\varepsilon_{Dr})$ of the complex dielectric constant for GaAs at the wavelength used in our pump-probe technique as a function of the thermalized electron concentration ($N_e^\mathrm{T}$). One can note that for wavelengths shorter than the intrinsic absorption edge of the semiconductor ($\lambda_g = 875$ nm for GaAs~\cite{sturge1963optical}), both the Drude model and the band filling effect contribute significantly to the modulation of the dielectric constant of the material. Conversely, in the infrared wavelength range, the dominant contribution is the generation of free carriers, described by the Drude model.

\begin{figure*}[h!]
    \centering
    \includegraphics[width=1\columnwidth]{Figs/supl_bf_dr}
    \caption{Dependence of the contributions of free carriers and the band filling effect in the modulation of real (blue lines) and imaginary(blue lines) parts of dielectric permittivity on the concentration of nonequilibrium thermalized electrons $N^{T}$. Graphs a) correspond to the pump radiation wavelength $\lambda_{\mathrm{pump}}= 500$ nm, and graphs b) are for fixed probe wavelength $\lambda_{\mathrm{pr}}= 1600$ nm. The dashed lines indicate the effect of free charge carriers according to the Drude model, and the solid lines are associated with the band filling effect.} 
     \label{fig_supl_bf_dr}
\end{figure*}

\newpage
\section{Numerical model}
To implement our theoretical approach and resolve the system of differential equations that describe the evolution of key characteristics self-consistently with Maxwell's equations governing sources within the system, we develop a numerical model using the COMSOL Multiphysics simulation package.

\begin{figure}[h!]
\center
\includegraphics[width=0.5\textwidth]{Figs/supl_bound.pdf}
\caption{Schematic representation of computational area and boundary conditions at various surfaces and boundaries in COMSOL Multiphysics model with the determination of initial conditions. Each color demonstrates the boundary and the equations defined on it. The scheme also shows in red the variables $T_e$, $T_{ph}$, $N_e^{\mathrm{T}}$, $N_e^{\mathrm{NT}}$ and $U_{e}^{\mathrm{NT}}$,  included in the differential equations and specified in certain computational domains.}
\label{fig_bound}
\end{figure}

The Electromagnetic Wave in Frequency Domain (EWFD) module, which utilizes the finite element method (FEM) for solving electromagnetic equations, is employed to simulate the optical response of the semiconductor film. When irradiated by two oppositely directed pumps, we utilize two separate EWFD modules corresponding to each irradiation.

Additionally, by linking the EWFD modules with the Partial Differential Equations (PDE) module, we can perform self-consistent calculations, taking into account the evolution of dielectric permittivity during the pulse. As a result, our model  consider the mutual influence between the distribution of the electromagnetic field $\mathbf{E}(\mathbf{r}, t)$ inside the semiconductor film and the evolution of crucial characteristics ($\mathcal{U}_{\mathrm{e}}^{ \mathrm{NT}}( \mathbf{r}, t),{N}_{\mathrm{ e}}^{\mathrm{NT}}(\mathbf{r}, t),{N}_{\mathrm{ e}}^{\mathrm{T}}(\mathbf{r}, t),{T}_{\mathrm{ e}}(\mathbf{r}, t),{T}_{\mathrm{ph}}(\mathbf{r}, t)$) in GaAs film.

In EWFD modules, incident pumps are defined through periodic ports, as illustrated in FIG.~\ref{fig_bound}, and periodic boundary conditions are applied with Floquet periodicity determined by these periodic ports.

For the key characteristics and corresponding PDE modules, we implement periodic boundary conditions (continuity) along the edges of the computational domain to simulate an infinite plane film. Additionally, to account for surface recombination of thermalized electrons, we introduce a surface flux term $q = -\gamma_{surf}N_e^{T}$, where $\gamma_{surf}$ represents the surface non-radiative relaxation coefficient at the interface between air and GaAs.

It is worth noting that the theoretical model by itself does not consider thermal interaction with the environment and subsequent cooling, as these processes occur over much longer time scales, typically on the order of nanoseconds. However, in our numerical calculations for the phonon temperature, we define $T_{ph}$ both in the GaAs film and in the Ag substrate. The latter allows us to take into account the heat outflow of heat into the substrate, based on the thermal properties of the Ag material, such as $C_{ph} = 234~~\mathrm{J/(kg\cdot K)}\cdot \rho$, $k_{ph} = 4.19~\mathrm{W/(cm\cdot K)}$, and $\mathrm{\rho = 0.0105~\mathrm{kg/cm^3}}$ - the lattice heat capacity, thermal conductivity, and density, respectively~\cite{BibEntry2023Nov}. The lower surface of the Ag substrate within the computational domain is determined by Dirichlet boundary conditions with $T=T_{ph}^0 = 300 $ K - the initial temperature of the lattice equal to room temperature. Additionally, we incorporate a convection boundary condition at the interface between the film and air $q = h_{air}(T_{ph}^0-T_{ph})$, where $h_{air} = 5~\mathrm{W/(m^2\cdot K)}$~\cite{Pokorny2010Jun}. This takes into account the cooling of the film through convection in the upper part of the film.  

On the contrary, all other key characteristics (except for $T_ph$) are determined solely within the GaAs film. Consequently, we apply Zero Flux boundary conditions on $\mathcal{U}_e^\mathrm{NT}$, $N_e^\mathrm{NT}$ and $T_e$ to restrict heat and particles' flow at the film surfaces.

Finally, the initial conditions for the electron concentration, as well as
the electron and phonon temperatures are the following $N_\mathrm{T}^0=2.1\cdot 10^{6}~\mathrm{cm^{-3}}$, $T_e^0 = T_{ph}^0 = 300~\mathrm{K}$. Moreover, the initial values of the energy $\mathcal{U}_\mathrm{NT}^\mathrm{0}$ and concentration ${N}_\mathrm{NT}^\mathrm{0}$ of non-thermalized electrons are assumed to be zero.

\section{Optically-induced diffraction grating leading to SPP excitation}
\subsection{Evolution of key characteristics}

Now, let's examine the evolution of key characteristics, focusing on the midpoint O within the lattice period as an illustrative example (FIG.~\ref{fig_supl_evolution}).

\begin{figure*}[h!]
    \centering
    \includegraphics[width=1\columnwidth]{Figs/supl_evolution}     \caption{
   The dynamics of the key characteristics for semiconductor material, such as non-thermalized electron concentration $N_e^\mathrm{NT}$, thermalized electron concentration $N_e^\mathrm{T}$ (b), as well as energy of non-thermalized electrons $\mathcal{U}_e^\mathrm{NT}$, electron temperature $T_{e}$ and phonon temperature $T_{ph}$ (a) at point O, illustrated in FIG.1 a) in the main text. Point O represents a location at the air/film interface positioned at the midpoint of the grating period. The Gaussian pump profile is shown as a dashed pink line.}  
     \label{fig_supl_evolution}
\end{figure*}

Initially, non-thermalized electrons are generated, causing their concentration ${N}_e^\mathrm{NT}$ to quickly increase from zero when the pulse starts. These electrons then rapidly transition into a thermalized state. As a result, after a short delay, the concentration of thermalized electrons begins to rise and eventually reaches its maximum (about 300 femtoseconds after the peak of the Gaussian pulse), and then begins to decrease due to relaxation processes (FIG.~\ref{fig_supl_evolution} b)).

Moreover, under the influence of laser radiation electrons absorb electromagnetic energy, leading to a sharp increase in the energy of non-thermalized electrons $\mathcal{U}_e^\mathrm{NT}$, demonstrated in FIG.~\ref{fig_supl_evolution} a). Subsequently, a significant portion of this energy is rapidly transferred to thermalized electrons. So, after reaching the maximum, the energy of non-thermalized electrons drops quite sharply due to the transfer of energy predominantly to electrons. Such effect is associated with the remarkably short electron thermalization time $\tau_e$, estimated to be around 200 femtoseconds. The evolution of the electron temperature $T_e$ is difficult to analyze because it strongly depends on the evolution of the electron conductivity $C_e(N_e^\mathrm{T})$, as well as on the energy transfer caused by electron-phonon scattering. The phonon temperature $T_{ph}$ begins to increase much later than the electron temperature due to the slow energy transfer from thermalized and non-thermalized electrons to phonons. Within a timeframe of approximately 5 to 20 nanoseconds, the temperatures of phonons $T_{ph}$ and electrons $T_e$ are practically equalized, which leads to the establishment of temperature equilibrium.

\subsection{Significance of model self-consistency}

It appears that for femtosecond pulses, considering self-consistent sources and time-varying dielectric constantwhen solving a system of differential equations is optional. Our calculations indicate that incorporating this self-consistency results in a minor correction (up to five percent) to the key characteristics of the semiconductor, leading to a weak alteration in the evolution of the dielectric constant, as depicted in the FIG.~\ref{fig_supl_self_nonself}. This phenomenon arises from the ultrashort pulse duration, when the rapid modulation of the dielectric constant has a negligible impact. However, with longer pulse durations, this effect becomes more pronounced. Hence, it is noteworthy that our theoretical model enables to analyze not only ultrashort pulses but also longer pulses, where self-consistency significantly influences the final dielectric permittivity.
\begin{figure*}[h!]
    \centering
    \includegraphics[width=1\columnwidth]{Figs/supl_self_nonself2}     \caption{
    Evolution of variations in the real and imaginary components of the dielectric constant at point O for pump and probe wavelengths of $\lambda_{\mathrm{pump}}= 500$ nm (a) and $\lambda_{\mathrm{probe}}=1600$ nm (b), respectively. The solid line corresponds to a self-consistent solution of a system of differential equations taking into account the change in dielectric constant affecting the sources, while the dotted line considers only the influence of evolving key parameters on the dielectric constant, neglecting modulation in the sources during the pulse.}  
     \label{fig_supl_self_nonself}
\end{figure*}
\subsection{SPP excitation}

FIG.~\ref{fig_supl_maps} illustrates the modulation of the complex dielectric permittivity $\mathrm{Re}~\varepsilon + i\mathrm{Im}~\varepsilon$ for both the probe wavelength (b) and pump wavelength (c) across a broad range of probe times $\mathrm{t_{pr}}$. Maps (a) correspond to the corresponding probe times and highlights the dynamics of surface plasmon polariton (SPP) dispersion, showcasing as a distinct dip in reflectance. This dispersion evolution attributed to a time-dependent optically induced diffraction grating, represented in b).
\begin{figure*}[h!]
    \centering
    \includegraphics[width=1\columnwidth]{Figs/supl_figures_maps2}
    \caption{a) The evolution of reflectance over probe time $\mathrm{t_{pr}}$ in terms of the normalized wave number $k_{||}/k_0 = sin(\theta_{\mathrm{pr}})$ and the normalized frequency $\omega/\omega_{p} = a/\lambda_{\mathrm{pr}}$, where $a$ is the diffraction grating period. These dynamics demonstrate that the SPP dispersion shifts toward the higher frequency range over time. Dotted black line refers to fixed probe wavelength $\lambda_{\mathrm{pr}}= 1600$ nm.  b) The distribution of thermalized and non-thermalized electrons concentrations, as well as complex dieletric permittivity for probe wavelegth $\lambda_{\mathrm{pr}}= 1600$ nm within the film at different probe times $\mathrm{t_{pr}}.$ c) The distribution of thermalized and non-thermalized electrons concentrations, as well as complex dieletric permittivity for pump wavelegth $\lambda_{\mathrm{pump}}= 500$ nm within the film at the same probe times $\mathrm{t_{pr}}.$}  
     \label{fig_supl_maps}
\end{figure*}
\bibliography{bib}